\documentclass[aps,prb,preprint,amsmath,floatfix]{revtex4}

\usepackage{dcolumn}
\usepackage{graphicx}

\renewcommand\vec[1]{{\mathbf #1}}
\newcommand\f[1]{\footnotemark[#1]}

\begin{document}

\title{Model for the on-site matrix elements of the tight-binding hamiltonian of a strained crystal: Application to silicon, germanium and their alloys}

\author{Y. M. Niquet}
\email{yniquet@cea.fr}
\affiliation{CEA, Institute for Nanosciences and Cryogenics (INAC), SP2M/L\_Sim, 38054 Grenoble Cedex 9, France}

\author{D. Rideau, C. Tavernier and H. Jaouen}
\affiliation{STMicroelectronics, 850 rue Jean Monnet, BP 16, F-38926 Crolles Cedex, France}

\author{X. Blase}
\affiliation{Institut N\'{e}el, CNRS and Universit\'{e} Joseph Fourier, BP 166, 38042 Grenoble Cedex 9, France}

\date{\today}

\begin{abstract}
We discuss a model for the on-site matrix elements of the $sp^3d^5s^*$ tight-binding hamiltonian of a strained diamond or zinc-blende crystal or nanostructure. This model features on-site, off-diagonal couplings between the $s$, $p$ and $d$ orbitals, and is able to reproduce the effects of arbitrary strains on the band energies and effective masses in the full Brillouin zone. It introduces only a few additional parameters and is free from any ambiguities that might arise from the definition of the macroscopic strains as a function of the atomic positions. We apply this model to silicon, germanium and their alloys as an illustration. In particular, we make a detailed comparison of tight-binding and {\it ab initio} data on strained Si, Ge and SiGe.
\end{abstract}

\maketitle

\section{Introduction}
\label{sectionIntroduction}

The oncoming limits of conventional downscaling of field-effect transistors have strengthened the need for innovative device architectures.\cite{ITRS} In this context, the use of mechanical strains has become an attractive solution to improve the electrical performances by enhancing the carrier mobility.\cite{Thompson04,Irie04} As a matter of fact, strain engineering techniques such as the growth of a contact etch stop layer (CESL),\cite{Payet06} or Si channels strained by SiGe source and drain extensions are now widely spread in the semiconductor industry. More generally, the electronic properties of strained Si$_{1-x}$Ge$_{x}$ layers grown on Si$_{1-y}$Ge$_{y}$ buffers are attracting much attention.\cite{Schaffler97} These heterostructures, that can be integrated into Si-based electronics and photonics, indeed offer the opportunity to tune the band gap of the active layer.

The modeling of the electrical properties of such devices requires a detailed description of the effects of strains on the band structure. Over the past decades, the {\it ab initio} methods such as the density functional theory\cite{Hohenberg64,Parr89} (DFT) have provided comprehensive information about the deformation potentials of semiconductors.\cite{Walle86,Cardona87,Walle89,Resta90,Wei99,Li06} However, such {\it ab initio} methods require heavy computational resources and are not, therefore, suitable for the calculation of the transport properties of large systems. For that reason, the physics and electronic device community is actively developing more efficient semi-empirical approaches, such as the $\vec{k}\cdot\vec{p}$,\cite{Bastard88,Rideau06} the empirical pseudopotential\cite{Chelikowsky76,Zunger95,Zunger99} or the tight-binding\cite{Slater54,Delerue05} (TB) methods, that can work out the electronic structure of strained semiconductors devices. Among these semi-empirical approaches, the TB method has long proved successful in predicting the electronic properties of semiconductor nanostructures such as nanocrystals or nanowires. The use of an atomic orbitals basis set with interactions limited to a few nearest neighbors indeed allows the calculation of the wave functions of million atom systems.\cite{Niquet08,Klimeck07} The TB method is also well suited to quantum transport calculations,\cite{Luisier06,Luisier07,Klimeck07,Svizhenko07,Lherbier08} and to the atomic scale description of, e.g., impurities\cite{Martins05,Diarra08} or electron-phonon coupling.\cite{Delerue01} In this respect, the first nearest neighbors $sp^3d^5s^*$ model is one of the most accurate and efficient TB description of semiconductor materials.\cite{Jancu98}

The effects of strains are accounted for in TB models through the bond length dependence of the nearest neighbor parameters $V_{\mu\nu}$ ($\mu$ and $\nu$ being two orbitals on different atoms), which is usually fitted to a power law:\cite{Harrison79,Harrison80}
\begin{equation}
V_{\mu\nu}(d)=V_{\mu\nu}(d_0)\left(\frac{d_0}{d}\right)^{n_{\mu\nu}},
\label{eqHarrison}
\end{equation}
where $d$ is the distance between the two atoms in the strained crystal and $d_0$ is the equilibrium distance. Although some hydrostatic and uniaxial deformation potentials can be reproduced that way,\cite{Priester88} much better accuracy can be achieved with the introduction of strain-dependent on-site parameters.\cite{Brey82,Tserbak93,Boykin02,Jancu07,Boykin07} Indeed, hydrostatic strain shifts the average potential\cite{noteavg} in the crystal, while uniaxial and shear strains split the $p$ or $d$ orbitals of a given atom. In their original $sp^3d^5s^*$ parametrization, Jancu {\it et al.}\cite{Jancu98} therefore introduced a term that lifts the degeneracy between the $d_{yz}$, $d_{xz}$, and $d_{xy}$ orbitals under uniaxial $\langle001\rangle$ strain. Jancu and Voisin later generalized this approach to uniaxial $\langle111\rangle$ strain.\cite{Jancu07} These hamiltonians, however, feature the macroscopic strains $\varepsilon_{\alpha\beta}$, whose expression as a function of the atomic positions (the basic input of the TB method) is not univocal. Boykin {\it et al.}\cite{Boykin02,Boykin07} therefore introduced position-dependent orbital energies in the $sp^3d^5s^*$ hamiltonian. They could reproduce that way the valence band deformation potentials $a_v$ and $b_v$, but did not really improve on $d_v$. This limitation is a consequence of the ``diagonal'' assumption made in that model. Uniaxial $\langle 111\rangle$ strain indeed leaves, for example, the $p_x$, $p_y$ and $p_z$ orbitals of a given atom equivalent. It however couples these orbitals off the diagonal of the hamiltonian.

In this paper, we discuss a model for the on-site matrix elements of the $sp^3d^5s^*$ TB model, based on an explicit expression for the crystal field, assuming that the total potential is the sum of central, atomic contributions.\cite{Chadi89, Mercer94} It features off-diagonal couplings between different orbitals, and is able to reproduce the effects of arbitrary strains on the band energies and effective masses at all relevant k-points. It only involves a few additional parameters, is fully consistent with the symmetries of the crystal, and is free from any ambiguity that might arise from the introduction of the macroscopic strains $\varepsilon_{\alpha\beta}$ in an atomistic description. We present this model in section \ref{sectionModel}, then discuss its properties in section \ref{sectionDiscussion}. Finally, We apply this model to silicon, germanium and their alloys, which are the most relevant materials for microelectronics, in section \ref{sectionApplications}. We provide detailed comparisons with {\it ab initio} data on strained Si, Ge and SiGe, and discuss two important problems: the increase of the longitudinal effective mass under shear strains (missing in previous TB models), and the description of random alloys.

\section{Model}
\label{sectionModel}

In this section, we introduce the model for the on-site matrix elements of the $sp^3d^5s^*$ tight-binding hamiltonian. For the sake of simplicity, we focus on a homogeneously strained diamond or zinc-blende crystal, the application to arbitrary strains and other crystal structures being straightforward. We assume that the total potential in the crystal is the sum of central, atomic contributions $\nu_1(|\vec{r}-\vec{R}_i|)$ (sublattice 1) and $\nu_2(|\vec{r}-\vec{R}_i|)$ (sublattice 2), $\vec{R}_i$ being the atomic positions. In a first nearest neighbor (NN) approximation, the potential experienced by the orbitals of atom $i$ on sublattice 1 is therefore:
\begin{equation}
\nu(\vec{r})=\nu_1(|\vec{r}-\vec{R}_i|)+\sum_j^{\rm NN}\nu_2(|\vec{r}-\vec{R}_j|).
\end{equation}
This potential shifts the energy of the orbitals and couples them one to each other in the strained crystal. In particular, $\nu(\vec{r})$ might lift the degeneracy between the $p$ or between the $d$ orbitals of the atom. Our model is actually based on a first-order expansion of the on-site matrix elements of the potential $\nu(\vec{r})$ as a function of the atomic positions. In the following, we calculate the on-site hamiltonian of the $p$ orbitals of sublattice 1 as an example (parapraph \ref{subsectionp}). We then discuss the application to other orbitals and crystal structures in paragraph \ref{subsectionothers}.

\subsection{Case of $p$ orbitals}
\label{subsectionp}

Let $p_i^x$, $p_i^y$ and $p_i^z$ be the $p$ orbitals of atom $i$, and:
\begin{subequations}
\begin{eqnarray}
V_1&=&\left\langle p_i^x\left|\nu_1(|\vec{r}-\vec{R}_i|)\right|p_i^x\right\rangle \\
V_2^\sigma(d_{ij})&=&\left\langle p_i^\sigma\left|\nu_2(|\vec{r}-\vec{R}_j|)\right|p_i^\sigma\right\rangle \\
V_2^\pi(d_{ij})&=&\left\langle p_i^\pi\left|\nu_2(|\vec{r}-\vec{R}_j|)\right|p_i^\pi\right\rangle,
\end{eqnarray}
\end{subequations}
where $p_i^\sigma$ and $p_i^\pi$ are the $p$ orbitals aligned ($\sigma$) or orthogonal ($\pi$) to the bond axis $\vec{R}_{ij}=\vec{R}_j-\vec{R}_i$. Slater-Koster relations easily yield:\cite{Slater54, Chadi89, Mercer94}
\begin{eqnarray}
V_x&=&\left\langle p_i^x\left|\nu\right|p_i^x\right\rangle \nonumber \\
&=&V_1+\sum_j^{\rm NN}V_2^\pi(d_{ij})+\sum_j^{\rm NN} l_{ij}^2\left[V_2^\sigma(d_{ij})-V_2^\pi(d_{ij})\right] \nonumber \\
&=&V_1+\frac{1}{3}\sum_j^{\rm NN}\left[V_2^\sigma(d_{ij})+2V_2^\pi(d_{ij})\right] \nonumber \\
&+&\sum_j^{\rm NN}\left[l_{ij}^2-\frac{1}{3}\right]\left[V_2^\sigma(d_{ij})-V_2^\pi(d_{ij})\right],
\end{eqnarray}
where $l_{\rm ij}=\vec{x}\cdot\vec{R}_{ij}/d_{ij}$ is the cosine director along $x$. This expression has been arranged so that the last (angular) term of the third line is zero in the unstrained material or under hydrostatic pressure (where $\sum_j^{\rm NN}l_{ij}^2=4/3$ whatever the orientation of the crystal with respect to the principal axes). We next expand $V_2^\sigma(d_{ij})$ and $V_2^\pi(d_{ij})$ in powers of $d_{ij}-d_0$:
\begin{subequations}
\begin{eqnarray}
V_2^\sigma(d_{ij})&=&V_2^\sigma(d_0)+\frac{3}{4}\alpha_p^\sigma\frac{d_{ij}-d_0}{d_0}+... \\
V_2^\pi(d_{ij})&=&V_2^\pi(d_0)+\frac{3}{4}\alpha_p^\pi\frac{d_{ij}-d_0}{d_0}+...
\end{eqnarray}
\end{subequations}
We hence get:
\begin{eqnarray}
V_x=V_1&+&\frac{1}{3}\sum_j^{\rm NN}\left[V_2^\sigma(d_0)+2V_2^\pi(d_0)\right] \nonumber \\
&+&\frac{3}{4}\alpha_p\sum_j^{\rm NN}\frac{d_{ij}-d_0}{d_0} \label{eqVp} \\
&+&\sum_j^{\rm NN}\left[\beta_p^{(0)}+\beta_p^{(1)}\frac{d_{ij}-d_0}{d_0}\right]\left[l_{ij}^2-\frac{1}{3}\right], \nonumber
\end{eqnarray}
where\cite{notealpha} $\alpha_p=(\alpha_p^\sigma+2\alpha_p^\pi)/3$, $\beta_p^{(0)}=V_2^\sigma(d_0)-V_2^\pi(d_0)$ and $\beta_p^{(1)}=3(\alpha_p^\sigma-\alpha_p^\pi)/4$. The first line of Eq. (\ref{eqVp}) is part of the unstrained $p$ orbital energy $E_p^0$. The second line is actually proportional (to first-order in the $d_{ij}$'s) to the hydrostatic strain, i.e. proportional to the relative variation of the volume $\Omega$ of the unit cell (also see paragraph \ref{sectionDiscussion}). We thus define for convenience:
\begin{equation}
\frac{\Delta\Omega}{\Omega_0}=\frac{\Omega-\Omega_0}{\Omega_0}=\frac{3}{4}\sum_j^{\rm NN}\frac{d_{ij}-d_0}{d_0}+{\cal O}\left(d_{ij}\right),
\label{eqdeltaomega}
\end{equation}
where $\Omega_0$ is the unstrained volume of the unit cell. The $p_x$ orbital energy therefore reads with these assumptions:
\begin{equation}
E_x=E_p^0+\alpha_p\frac{\Delta\Omega}{\Omega_0}+\sum_j^{\rm NN}\beta_p(d_{ij})\left[l_{ij}^2-\frac{1}{3}\right],
\end{equation}
where $\beta_p(d)=\beta_p^{(0)}+\beta_p^{(1)}(d-d_0)/d_0$.

The equations are similar for $E_y$ and $E_z$, with $l_{ij}$ replaced by $m_{ij}=\vec{y}\cdot\vec{R}_{ij}/d_{ij}$ and $n_{ij}=\vec{z}\cdot\vec{R}_{ij}/d_{ij}$ respectively. Off-diagonal couplings between the $p$ orbitals can be obtained in the same way. Slater-Koster relations\cite{Slater54} yield for example:
\begin{eqnarray}
\left\langle p_i^y\left|\nu\right|p_i^x\right\rangle&=&\sum_j^{\rm NN}m_{ij}l_{ij}\left[V_2^\sigma(d_{ij})-V_2^\pi(d_{ij})\right] \nonumber \\
&=&\sum_j^{\rm NN}\beta_p(d_{ij})m_{ij}l_{ij},
\end{eqnarray}
which is also zero under hydrostatic pressure.

The on-site, $p$ block matrix finally reads in the $\{p_x,p_y,p_z\}$ basis set:
\begin{eqnarray}
\hat{\vec{H}}_p&=&\left(E_p^0+\alpha_p\frac{\Delta\Omega}{\Omega_0}\right)\hat{\vec{I}} \nonumber \\
&+&\sum_j^{\rm NN}\beta_p(d)
\begin{bmatrix}
l^2-\frac{1}{3} & lm & ln \\
ml & m^2-\frac{1}{3} & mn \\
nl & nm & n^2-\frac{1}{3} \\
\end{bmatrix},
\label{eqporbs}
\end{eqnarray}
where the explicit dependence of $d_{ij}$, $l_{ij}$, $m_{ij}$ and $n_{ij}$ on the atomic sites $i$ and $j$ has been dropped for simplicity. The $p$ orbitals feature a $\propto\alpha_p$ hydrostatic correction and a $\propto\beta_p$ angular term, whose effects will be discussed in more detail in section \ref{sectionDiscussion}.

\subsection{Case of other orbitals}
\label{subsectionothers}

The on-site hamiltonians of the $s$ ($s^*$) and $d$ orbitals, as well as the off-diagonal coupling matrices between the $s$, $p$, $d$ and $s^*$ orbitals are given in appendix \ref{AppendixCouplings}. Eq. (\ref{eqporbs}) as well as Eqs. (\ref{eqsorbs})--(\ref{eqpdorbs}) of appendix \ref{AppendixCouplings} are valid for both sublattices 1 and 2, possibly with different parameters in III-V or II-VI materials. They feature hydrostatic ($\propto\alpha$ terms) and/or angular terms ($\propto\beta$ and $\propto\gamma$ matrices). The $\propto\beta$ matrices are all zero in the unstrained crystal and under hydrostatic strain. There are, however, non-zero couplings between the $d$ orbitals [Eq. (\ref{eqdorbs})], between the $s$ and $s^*$ orbitals [Eq. (\ref{eqstorbs})], and between the $p$ and $\{d_{yz},d_{xz},d_{xy}\}$ orbitals [Eq. (\ref{eqpdorbs})] if the corresponding $\gamma$ parameters are not zero. In particular, the $d$ orbitals are not degenerate any more in the unstrained crystal if $\gamma_d^{(0)}\ne0$ (see paragraph \ref{subsectiond} of the appendix). This is actually consistent with the symmetry of the zinc-blende lattice, but is not, usually, accounted for in TB models. As a matter of fact, lifting the degeneracy between the $d$ orbitals does not significantly improve the quality of the TB model in diamond or zinc-blende crystals, while it is essential in lower symmetry polytypes such as wurtzite materials.

This model has been checked against an {\it ab initio} (DFT) description of silicon based on atomic-like orbitals (the SIESTA code\cite{SIESTA}).
With the single-$\zeta$-polarized basis set used, the self-consistent {\it ab initio} Hamiltonian is formally equivalent to a non-orthogonal third nearest neighbors $sp^3d^5$ TB model. The evolution of the on-site {\it ab initio} matrix elements under strain compares fairly well with our tight-binding approach (despite the latter being first nearest neighbors only). All $\beta^{(0)}$'s and $\gamma^{(0)}$'s (except $\beta_d^{(0)}$) are found negative within SIESTA, as expected from simple arguments assuming positive, exponentially decaying radial parts for the orbitals. The sign of the $\alpha$'s, $\beta^{(1)}$'s and $\gamma^{(1)}$'s is, however, expected to be quite sensitive to the choice of orbitals.\cite{notealpha, notebeta}

The present model can be applied to other crystal structures and inhomogeneous strains. In a wurtzite material for example, the $\propto\beta_p$ and $\propto\beta_d$ or $\gamma_d$ terms will lift the degeneracy between the $p$ and between the $d$ orbitals in the unstrained crystal, as is usually enforced {\it a priori} in the TB descriptions of these materials.\cite{Jancu02} We will now discuss some properties of this model, then its application to silicon, germanium and their alloys.

\section{Discussion}
\label{sectionDiscussion}

Eq. (\ref{eqporbs}) and Eqs. (\ref{eqsorbs})--(\ref{eqpdorbs}) directly depend on the atomic coordinates through the interatomic distances $d_{ij}$ and cosine directors $l_{ij}$, $m_{ij}$, and $n_{ij}$. These equations are thus free of any amibguities that might arise, e.g., from the definition of the strains $\varepsilon_{\alpha\beta}$ as a function of the atomic positions, in particular in inhomogeneous environments like alloys. They also account for internal strains at the atomistic level, and should therefore be able to reproduce electron-optical phonon couplings. Moreover, this model for the on-site tight-binding hamiltonian is consistent with the symmetries of the crystal. In particular, the band structure remains invariant under global rotation of the lattice (since these equations fulfill Slater-Koster's relations\cite{Slater54}), a property which is not easily enforced in models depending explicitely on the $\varepsilon_{\alpha\beta}$'s or in the model of Refs. \onlinecite{Boykin02} and \onlinecite{Boykin07}. In practice, the input atomic positions can be calculated using, for example, Keating's\cite{Keating66} or Stillinger-Weber force fields.\cite{Stillinger85}

We next discuss the effects of biaxial stress on the $p$ orbitals as an illustration of the versatility of this model. In a homogeneously strained crystal, the strained atomic positions $\vec{R}_i$ read as a function of the unstrained coordinates $\vec{R}_i^0$:
\begin{equation}
\vec{R}_i=(\hat{\vec{I}}+\hat{\vec{\varepsilon}})\vec{R}_i^0\pm\zeta\frac{a}{4}(\varepsilon_{yz},\varepsilon_{xz},\varepsilon_{xy}),
\label{eqRi}
\end{equation}
where the $+$ (resp. $-$) sign holds for sublattice 1 (resp. sublattice 2), $\zeta$ is Kleinman's internal strain parameter, $a$ is the lattice parameter, $\hat{\vec{I}}$ is the identity matrix and $\hat{\vec{\varepsilon}}$ is the matrix of the strains $\varepsilon_{\alpha\beta}$. The internal strain parameter $\zeta$ describes the motion of one sublattice with respect to the other under shear strain.\cite{Kleinman62} We successively consider the cases of biaxial $\langle001\rangle$ and $\langle111\rangle$ strains.

\subsection{The case of biaxial $\langle001\rangle$ strain}
\label{subsection001}

Let us apply a biaxial stress perpendicular to $z=[001]$, and let $\varepsilon_{xx}=\varepsilon_{yy}=\varepsilon_{\parallel}$ and $\varepsilon_{zz}=\varepsilon_{\perp}$ be the strains in the crystal. Eqs. (\ref{eqporbs}) and (\ref{eqRi}) then yield, to first-order in strains:
\begin{eqnarray}
\hat{\vec{H}}_p&=&\left(E_p^0+\alpha_p\frac{\delta\Omega}{\Omega_0}\right)\hat{\vec{I}} \nonumber \\
&+&\frac{8}{9}\beta_p^{(0)}(\varepsilon_{\perp}-\varepsilon_{\parallel})
\begin{bmatrix}
-1 &  0 & 0 \\
 0 & -1 & 0 \\
 0 &  0 & 2 \\
\end{bmatrix}.
\end{eqnarray}
The first line features the hydrostatic strain $\delta\Omega/\Omega_0=\varepsilon_{xx}+\varepsilon_{yy}+\varepsilon_{zz}=2\varepsilon_{\parallel}+\varepsilon_{\perp}$. It accounts for the variation of the average potential in the crystal and shifts the three $p$ orbitals equally. As expected, the stress also lifts (second line) the degeneracy between the $\{p_x, p_y\}$ and the $p_z$ orbitals. The splitting between $\{p_x, p_y\}$ and $p_z$, $\delta E_p=8\beta_p^{(0)}(\varepsilon_{\perp}-\varepsilon_{\parallel})/3$, is actually proportional to the uniaxial component of the strain tensor, but does not depend on $\beta_p^{(1)}$. The degeneracy between $\{d_{yz},d_{xz}\}$ and $d_{xy}$ is likewise lifted for the $d$ orbitals. This model thus reproduces the effects of the $\propto\delta_{001}$ term in the parametrizations of Jancu {\it et al.},\cite{Jancu98,Jancu07} or of the diagonal energy shifts in the parametrization of Boykin {\it et al}.\cite{Boykin02,Boykin07}

\subsection{The case of biaxial $\langle111\rangle$ strain}
\label{subsection111}

Let us now apply a biaxial stress perpendicular to $z'=[111]$. The strains in the $\{x'=[1\bar10],y'=[11\bar2],z'=[111]\}$ axis set are thus $\varepsilon_{x'x'}=\varepsilon_{y'y'}=\varepsilon_{\parallel}$ and $\varepsilon_{z'z'}=\varepsilon_{\perp}$. Eqs. (\ref{eqporbs}) and (\ref{eqRi}) then yield, to first-order in strains:
\begin{equation}
\hat{\vec{H}}_p=\left(E_p^0+\alpha_p\frac{\delta\Omega}{\Omega_0}\right)\hat{\vec{I}}+\frac{8}{27}\beta_p^{\rm eff}(\varepsilon_{\perp}-\varepsilon_{\parallel})
\begin{bmatrix}
0 & 1 & 1 \\
1 & 0 & 1 \\
1 & 1 & 0 \\
\end{bmatrix},
\label{eqp111}
\end{equation}
where $\beta_p^{\rm eff}=\beta_p^{(0)}(1+2\zeta)+\beta_p^{(1)}(1-\zeta)$. As expected, biaxial $[111]$ strain leaves the $p_x$, $p_y$ and $p_z$ (diagonal) energies equivalent. It however couples these orbitals off the diagonal of the hamiltonian. The eigenvectors of $\hat{\vec{H}}_p$ are indeed: {\it i}) the $p$ orbital aligned with $[111]$ ($p_{z'}$), with energy $E_p^0+\alpha_p\delta\Omega/\Omega_0+16\beta_p^{\rm eff}(\varepsilon_{\perp}-\varepsilon_{\parallel})/27$, and {\it ii}) the two degenerate $p$ orbitals perpendicular to $[111]$ ($\{p_{x'}, p_{y'}\}$), with energies $E_p^0+\alpha_p\delta\Omega/\Omega_0-8\beta_p^{\rm eff}(\varepsilon_{\perp}-\varepsilon_{\parallel})/27$. The splitting between these orbitals is again proportional to the uniaxial component of the strain tensor. It also depends on the internal strain parameter $\zeta$ (through $\beta_p^{\rm eff}$). The value of $\zeta$ used as a reference to compute the deformation potentials must therefore be provided with the TB parameters.

Such off-diagonal couplings between the $p$ (or $d$) orbitals do not exist in the parametrization of Ref. \onlinecite{Boykin02}. As a consequence the degeneracy between the $p$ and between the $d$ orbitals is not lifted by biaxial $\langle 111\rangle$ strain, and the value of $d_v$ is the same whether the diagonal energy corrections are included or not. $\beta_p^{(1)}$ and $\beta_d^{(1)}$ also reproduce the effects of the $\propto\delta_{111}$ and $\propto\pi_{111}$ terms in the parametrization of Ref. \onlinecite{Jancu07}. However, the effective $\beta_d^{(0)}$ is assumed to be zero for $[111]$ strain (but not for $[001]$ strain), which makes the model of Ref. \onlinecite{Jancu07} hardly consistent with an explicit description of the crystal field, even beyond first nearest neighbors.

\section{Application to Si, Ge and their alloys}
\label{sectionApplications}

In this section, we discuss the application of the above model for the on-site matrix elements of the TB hamiltonian to silicium, germanium and their alloys. We therefore attempted to reproduce the band structure of Si, Ge and of the ordered Si$_{0.5}$Ge$_{0.5}$ alloy with a first nearest neighbor, two-center orthogonal $sp^3d^5s^*$ TB model. We used experimental data when available and {\it ab initio} calculations otherwise as a reference for the optimization of the TB parameters. We first review the {\it ab initio} calculations and the optimization process in paragraph \ref{subsectionabinitio}, then discuss the TB model of Si, Ge and Si$_{0.5}$Ge$_{0.5}$ in paragraph \ref{subsectionoresultsSiGe}, and finally the case of arbitrary SiGe alloys in paragraph \ref{subsectionalloys}.

\subsection{First principle calculations and optimization procedure}
\label{subsectionabinitio}

A series of first principle calculations was performed with the ABINIT\cite{abinit,abinit02,abinit05} code on Si, Ge, and the ordered Si$_{0.5}$Ge$_{0.5}$ alloy, to set up a reference for the optimization of the TB parameters. These calculations are based on the local density approximation (LDA) to DFT,\cite{Hohenberg64,Parr89} using relativistic Hartwigsen-Goedecker-Hutter pseudo-potentials.\cite{Hartwigsen98} The LDA band structure was further corrected with Hedin's GW approximation to the self-energy used as a post-DFT scheme.\cite{Hedin65,Aulbur00} In general, the GW band energies were found in good agreement with the available experimental data.\cite{detailonGW} The properties of a large set of strained crystals have been computed, including hydrostatic as well as biaxial deformations perpendicular to [100], [110], and [111].\cite{detailonGWstr} The biaxial strains were chosen large enough (up to $\varepsilon_\parallel=\pm5\%$) to span the whole range of lattice mismatches encountered in epitaxial Si$_{1-x}$Ge$_x$ layers grown on relaxed Si$_{1-y}$Ge$_y$ buffers. The atomic positions within the cell were carefully optimized, as they strongly affect the band structure.\cite{Rideau06}

The TB parameters were fitted to the {\it ab initio} (or experimental, when available) band structures, effective masses and deformation potentials using global optimization methods\cite{Jones93} refined with local optimizers.\cite{Byrd93} The least-square convergence of the band structures was monitored on a dense set of $k$-points in the first Brillouin zone.

The $sp^3d^5s^*$ model of Si and Ge features 4 on-site energies and $\alpha$ parameters, 14 nearest neighbor and Harrison ($n_{\mu\nu}$) parameters, and up to 20 $\beta$ and $\gamma$ parameters. However, only 9 of them appeared to have significant impact on the electronic structure of strained Si and Ge around the band gap (see Table \ref{tableparamsSiGestrains} for a list). In particular, all $\gamma$ parameters and most $\beta^{(1)}$'s were set to zero. This left 45 parameters in the model, that were optimized in following way:
\begin{enumerate}
\item The 4 on-site energies and 14 nearest neighbors parameters were fitted on the band structures of relaxed Si and Ge.
\item The 4 $\alpha$'s and 14 Harrison parameters were fitted on one positive and one negative hydrostatic strain.\cite{notefithydro}
\item The 7 $\beta^{(0)}$'s were fitted on one $[100]$ and one $[111]$ biaxial strain that do not change the first nearest neighbor bond lengths ($\varepsilon_\perp\simeq-2\varepsilon_\parallel$ and $\zeta\simeq1$).
\item The 14 Harrison, 7 $\beta^{(0)}$'s and 2 $\beta^{(1)}$'s were further refined on one $[100]$ and two $[111]$ biaxial strains [one with $\zeta=0.557$ (Si) or $\zeta=0.536$ (Ge) and one with $\zeta=0$].
\end{enumerate}
Steps 2 and 3 ensure a reasonnable starting point for step 4. The resulting parametrization was also checked against $[110]$ biaxial strains, and its transferability tested on strained Si/Ge films and wires.

The TB model of the Si$_{0.5}$Ge$_{0.5}$ alloy only involves 7 additional first nearest neighbor parameters (since the Si/Ge and Ge/Si interactions are different). The on-site energies and on-site strain parameters of the Si and Ge atoms were chosen equal to those of bulk Si and Ge respectively.
%The average of the Harrison parameters of bulk Si and Ge turned out to yield an excellent description of strained Si$_{0.5}$Ge$_{0.5}$, as discussed in next paragraph.

The TB parameters of Si, Ge and Si$_{0.5}$Ge$_{0.5}$ are listed in Tables \ref{tableparamsSiGe}, \ref{tableparamsSiGestrains}, and \ref{tableparamsSiGealloy}. The on-site strain parameters have the sign expected from simple considerations about the shape of the orbitals, except $\beta_p^{(0)}$, $\beta_{sp}^{(0)}$ and $\beta_{sd}^{(0)}$. We point out that the sign of these three parameters is extremely robust; Including the missing $\gamma^{(0)}$'s in the on-site corrections will not, in particular, change the picture.\cite{notegamma0} The positive sign of $\beta_p^{(0)}$  seems characteristic of first nearest neighbors orthogonal models: The model of Ref. \onlinecite{Boykin07} indeed splits the $p$ orbitals the same way as ours; while this is hidden in Ref. \onlinecite{Jancu07} by the choice of an effective $\beta_p^{(0)}=0$ but different effective $\beta_d^{(0)}$'s for biaxial $[001]$ and $[111]$ strains (see discussion in paragraph \ref{subsection111}). The reasons are twofold: First, the orbitals hidden behind orthogonal TB models are much more complex than usually assumed when discussing the sign of the interactions. The radial parts must indeed have at least one zero to fulfill (near) orthogonality relations with the neighboring atoms. Second, some deformation potentials, such as $b_v$ and $\Xi_u^\Delta$, are independent on the first (but not on the second) nearest neighbor Harisson parameters ($b_v$, for example depends, on $\beta_p^{(0)}$ and $\beta_d^{(0)}$ only). The on-site parameters will therefore likely renormalize the missing long-range interactions beyond their ``bare'' definition given in section \ref{subsectionp} and appendix \ref{AppendixCouplings}. The renormalization of long-range interactions into first nearest neighbor and on-site terms is underlying every short-range TB model and is a key of their success. We have carefully checked our parametrization in bulk (including properties that were not included in the optimization, such as the non linearities of the band edges and the behavior of the masses under shear strains discussed in the next paragraph), and tested its transferability to random SiGe alloys (paragraph \ref{subsectionalloys}) and to a variety of test nanostructures such as strained Si/Ge films and wires. This model (as the previous ones) actually shows excellent transferability of the bulk physics to the nanostructures.

\subsection{Results in bulk Si, Ge, and SiGe}
\label{subsectionoresultsSiGe}

\begin{table}
\centering
%\newcolumntype{d}[1]{D{.}{.}{#1}}
\begin{tabular}{lddl}
\toprule
 & \multicolumn{1}{c}{Si} & \multicolumn{1}{c}{Ge} & \\
\hline
$E_s$ & -2.55247 & -4.08253 & eV \\
$E_p$ & 4.48593 & 4.63470 & eV \\
$E_d$ & 14.01053 & 12.19526 & eV \\
$E_{s^*}$ & 23.44607 & 23.20167 & eV \\
$\lambda_{\rm so}$ & 0.01851 & 0.12742 & eV \\
\hline
$V_{ss\sigma}$ & -1.86600 & -1.49093 & eV \\
$V_{ss^*\sigma}$ & -1.39107 & -1.59479 & eV \\
$V_{sp\sigma}$ & 2.91067 & 2.91277 & eV \\
$V_{sd\sigma}$ & -2.23992 & -2.10114 & eV \\
$V_{s^*s^*\sigma}$ & -4.51331 & -4.86118 & eV \\
$V_{s^*p\sigma}$ & 3.06822 & 2.92036 & eV \\
$V_{s^*d\sigma}$ & -0.77711 & -0.23561 & eV \\
$V_{pp\sigma}$ & 4.08481 & 4.36624 & eV \\
$V_{pp\pi}$ & -1.49207 & -1.58305 & eV \\
$V_{pd\sigma}$ & -1.66657 & -1.60110 & eV \\
$V_{pd\pi}$ & 2.39936 & 2.36977 & eV \\
$V_{dd\sigma}$ & -1.82945 & -1.15483 & eV \\
$V_{dd\pi}$ & 3.08177 & 2.30042 & eV \\
$V_{dd\delta}$ & -1.56676 & -1.19386 & eV \\
\botrule
\end{tabular}
\caption{Tight-binding parameters of relaxed, bulk Si and Ge (first nearest neighbor, two-center orthogonal $sp^3d^5s^*$ model). The notations are those of Slater and Koster.\cite{Slater54} The valence bands of Si and Ge have been aligned at $E=0$ eV; The on-site energies $E_s$, $E_p$, $E_d$, and $E_{s^*}$ of Ge must, therefore, be shifted by $\Delta_{\rm VBO}=0.68$ eV to account for the valence band offset between the two materials. $\lambda_{\rm so}$ is the spin-orbit coupling parameter of the $p$ orbitals.}
\label{tableparamsSiGe}
\end{table}

\begin{table}
\centering
%\newcolumntype{d}[1]{D{.}{.}{#1}}
\begin{tabular}{lddl}
\toprule
 & \multicolumn{1}{c}{Si} & \multicolumn{1}{c}{Ge} & \\
\hline
$d_0$ & 2.35169 & 2.44999 & \AA  \\
\hline
$n_{ss\sigma}$ & 3.56701 & 3.57536 &    \\
$n_{ss^*\sigma}$ & 1.51967 & 1.03634 &    \\
$n_{sp\sigma}$ & 2.03530 & 2.88203 &    \\
$n_{sd\sigma}$ & 2.14811 & 1.89283 &    \\
$n_{s^*s^*\sigma}$ & 0.64401 & 1.07935 &    \\
$n_{s^*p\sigma}$ & 1.46652 & 2.64809 &    \\
$n_{s^*d\sigma}$ & 1.79667 & 2.33424 &    \\
$n_{pp\sigma}$ & 2.01907 & 2.40576 &    \\
$n_{pp\pi}$ & 2.87276 & 2.95026 &    \\
$n_{pd\sigma}$ & 1.00446 & 0.51325 &    \\
$n_{pd\pi}$ & 1.78029 & 1.62421 &    \\
$n_{dd\sigma}$ & 1.73865 & 1.68410 &    \\
$n_{dd\pi}$ & 1.80442 & 2.64952 &    \\
$n_{dd\delta}$ & 2.54691 & 3.83221 &    \\
\hline
$\alpha_s$ & -0.13357 & -0.33252 & eV \\
$\alpha_p$ & -0.18953 & -0.43824 & eV \\
$\alpha_d$ & -0.89046 & -0.90486 & eV \\
$\alpha_{s^*}$ & -0.24373 & -0.52062 & eV \\
$\beta_p^{(0)}$ & 1.13646 & 1.01233 & eV \\
$\beta_p^{(1)}$ & -2.76257 & -2.53951 & eV \\
$\beta_{pd}^{(0)}$ & -0.13011 & -0.22597 & eV \\
$\beta_{pd}^{(1)}$ & -3.28537 & -3.77180 & eV \\
$\beta_d^{(0)}$ & 3.59603 & 1.99217 & eV \\
$\beta_{sp}^{(0)}$ & 1.97665 & 1.27627 & eV \\
$\beta_{s^*p}^{(0)}$ & -2.18403 & -2.02374 & eV \\
$\beta_{sd}^{(0)}$ & 3.06840 & 2.38822 & eV \\
$\beta_{s^*d}^{(0)}$ & -4.95860 & -4.73191 & eV \\
\botrule
\end{tabular}
\caption{Harrison and on-site strain parameters of Si and Ge.}
\label{tableparamsSiGestrains}
\end{table}

\begin{table}
\centering
%\newcolumntype{d}[1]{D{.}{.}{#1}}
\begin{tabular}{ldlldlld}
\toprule
\multicolumn{8}{c}{Si(1)Ge(2)} \\
\hline
$V_{s_1s_2\sigma}$ & -1.67650 & eV & & & & $n_{ss\sigma}$ & 3.90172 \\
$V_{s_1^*s_2^*\sigma}$ & -4.63349 & eV & & & & $n_{s^*s^*\sigma}$ & 0.85993 \\
$V_{p_1p_2\sigma}$ & 4.21933 & eV & & & & $n_{pp\sigma}$ & 2.34995 \\
$V_{p_1p_2\pi}$ & -1.54668 & eV & & & & $n_{pp\pi}$ & 3.08150 \\
$V_{d_1d_2\sigma}$ & -1.41949 & eV & & & & $n_{dd\sigma}$ & 1.66975 \\
$V_{d_1d_2\pi}$ & 2.62540 & eV & & & & $n_{dd\pi}$ & 2.24973 \\
$V_{d_1d_2\delta}$ & -1.39382 & eV & & & & $n_{dd\delta}$ & 3.06305 \\
$V_{s_1s_2^*\sigma}$ & -1.50940 & eV & $V_{s_2s_1^*\sigma}$ & -1.50314 & eV & $n_{ss^*\sigma}$ & 1.03801 \\
$V_{s_1p_2\sigma}$ & 2.82890 & eV & $V_{s_2p_1\sigma}$ & 3.01033 & eV & $n_{sp\sigma}$ & 2.37280 \\
$V_{s_1d_2\sigma}$ & -2.13989 & eV & $V_{s_2d_1\sigma}$ & -2.04737 & eV & $n_{sd\sigma}$ & 1.99537 \\
$V_{s_1^*p_2\sigma}$ & 3.06299 & eV & $V_{s_1^*p_2\sigma}$ & 2.79296 & eV & $n_{s^*p\sigma}$ & 1.94143 \\
$V_{s_1^*d_2\sigma}$ & -0.46386 & eV & $V_{s_2^*d_1\sigma}$ & -0.51235 & eV & $n_{s^*d\sigma}$ & 2.01051 \\
$V_{p_1d_2\sigma}$ & -1.43412 & eV & $V_{p_2d_1\sigma}$ & -1.61322 & eV & $n_{pd\sigma}$ & 0.75549 \\
$V_{p_1d_2\pi}$ & 2.57110 & eV & $V_{p_2d_1\pi}$ & 2.43552 & eV & $n_{pd\pi}$ & 1.67031 \\
\botrule
\end{tabular}
\caption{First nearest neighbor two-center tight-binding parameters of SiGe. The on-site energies and on-site strain parameters of the Si and Ge atoms are those of bulk Si and Ge respectively. The Harrison parameters are the same for Si/Ge and Ge/Si interactions. The relaxed SiGe bond length is $d_0=2.39792$ \AA.}
\label{tableparamsSiGealloy}
\end{table}

\begin{figure}
\centering
\includegraphics[width=0.66\columnwidth]{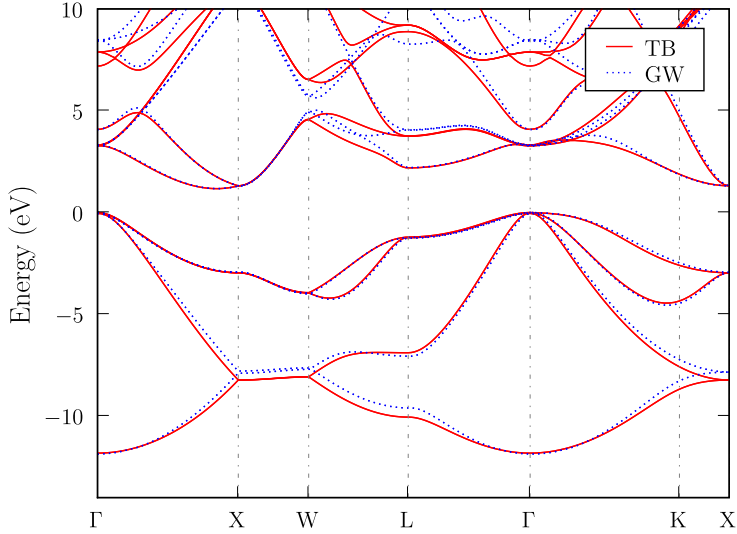}
\caption{(Color online) Band structure of bulk, unstrained silicon in the $sp^3d^5s^*$ TB and GW approximations.}
\label{figBSSi}
\end{figure}

\begin{figure}
\centering
\includegraphics[width=0.66\columnwidth]{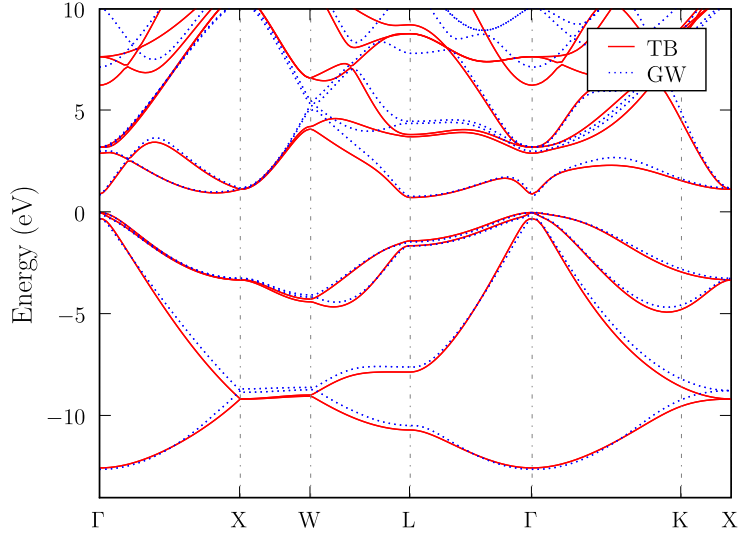}
\caption{(Color online) Band structure of bulk, unstrained germanium in the $sp^3d^5s^*$ TB and GW approximations.}
\label{figBSGe}
\end{figure}

\begin{figure}
\centering
\includegraphics[width=0.66\columnwidth]{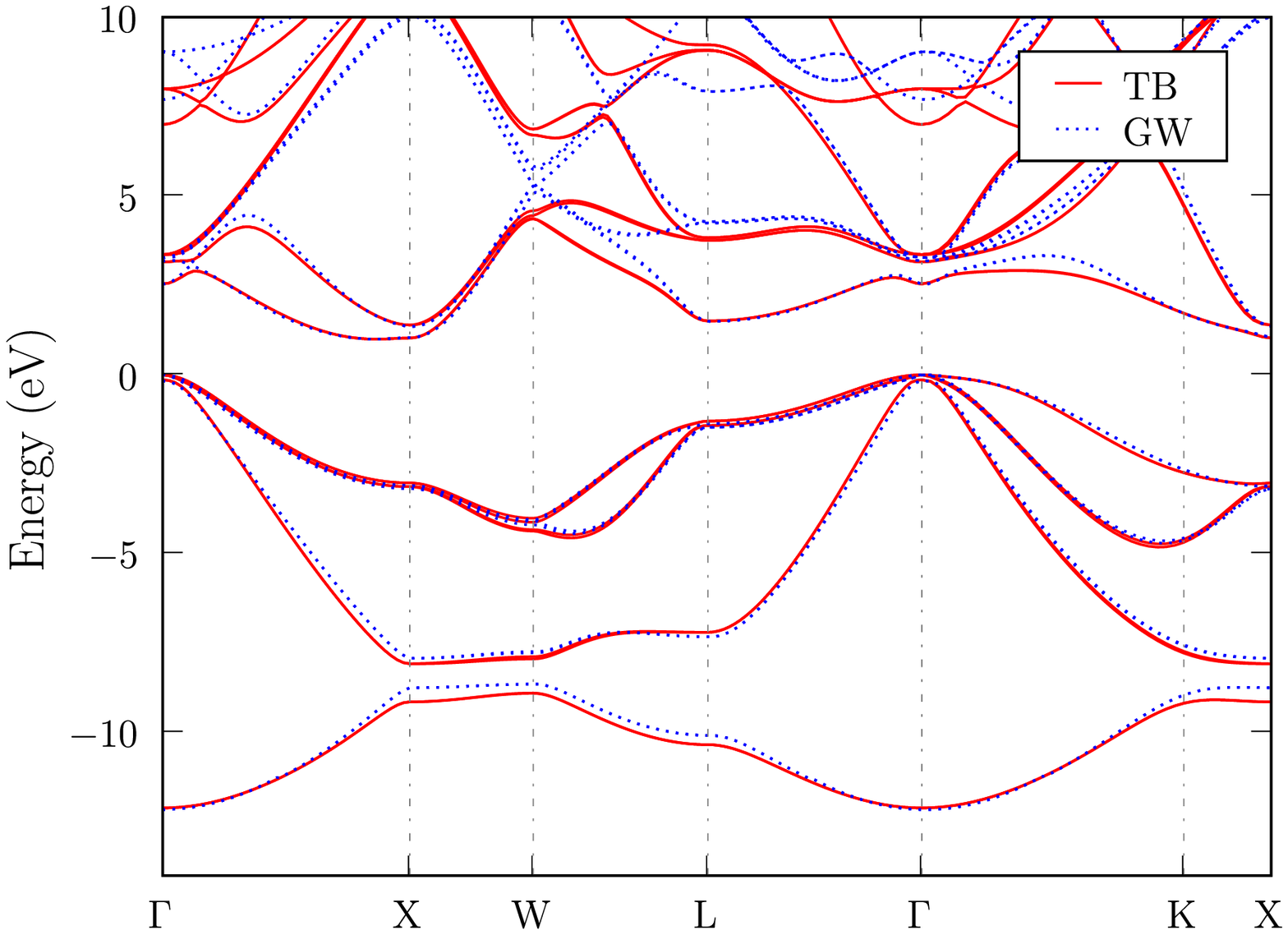}
\caption{(Color online) Band structure of bulk, unstrained Si$_{0.5}$Ge$_{0.5}$ in the $sp^3d^5s^*$ TB and GW approximations.}
\label{figBSSiGe}
\end{figure}

The TB and GW band structures of bulk, unstrained Si, Ge and Si$_{0.5}$Ge$_{0.5}$ are compared in Figs. \ref{figBSSi}, \ref{figBSGe}, and \ref{figBSSiGe}. They are in very good agreement one with each other, the difference between the TB and GW principal band gaps being $<0.01$ eV. The lifting of the degeneracies at, e.g., the X point in SiGe are also well reproduced. The TB conduction band effective masses and valence band Luttinger parameters of Si and Ge are given in Table \ref{tablemasses}. They are compared with the GW and experimental data, and with two other $sp^3d^5s^*$ parameterizations.\cite{Jancu98,Boykin04} %The present model is of comparable, or even sligthy better accuracy (especially for Si) than these previous parametrizations. %Similar agreement was found, e.g., on the longitudinal effective mass of the $\Delta$ valleys in Si$_{0.5}$Ge$_{0.5}$ ($m_l^\Delta=1.12$ instead of $m_l^\Delta=1.16$).

% \begin{table}
% \centering
% %\newcolumntype{d}[1]{D{.}{.}{#1}}
% \begin{ruledtabular}
% \begin{tabular}{lddd}
% Si & \multicolumn{1}{c}{Exp.\f{1}} & \multicolumn{1}{c}{GW\f{2}} & \multicolumn{1}{c}{TB}\\
% \hline
% $E_g(\Delta)$ & 1.170 & (1.170) & 1.171  \\
% 
% Ge & \multicolumn{1}{c}{Exp.\f{1}} & \multicolumn{1}{c}{GW\f{2}} & \multicolumn{1}{c}{TB}\\
% \hline
% $E_g({\rm L})$& 0.744 & (0.744) & 0.738  \\
% $E_g(\Gamma) $& 0.900 &  0.000  & 0.892  \\
% $E_g(\Delta) $&       &  0.959  & 0.964  \\
% \end{tabular}
% \end{ruledtabular}
% \footnotetext[1]{Ref.~\onlinecite{Landolt}.}
% \footnotetext[2]{See Ref.~\onlinecite{detailonGW} for details about the GW band gap.}
% \caption{Band gap energies of Si and Ge (eV).}
% \label{tablegaps}
% \end{table}

\begin{table}
\centering
\begin{ruledtabular}
\begin{tabular}{lcccc}
Si & Exp. & GW & $sp^3d^5s^*$ TB & Present model\\
\hline
$m_l^\Delta $&$0.9163\f{1}$&$0.925$&$0.702\f{8}; 0.891\f{9}$&$0.900$ \\
$m_t^\Delta $&$0.1905\f{1}$&$0.189$&$0.227\f{8}; 0.201\f{9}$&$0.197$ \\
$m_l^{\rm L}$&$           $&$1.808$&$1.378\f{8}; 3.433\f{9}$&$2.125$ \\
$m_t^{\rm L}$&$           $&$0.124$&$0.161\f{8}; 0.174\f{9}$&$0.151$ \\

$\gamma_1   $&$4.26\f{1}; 4.285\f{1}$&$4.54$&$4.51\f{8}; 4.15\f{9}$&$4.22$ \\
             &$4.22\f{2}; 4.340\f{3}$ \\
$\gamma_2   $&$0.38\f{1}; 0.339\f{1}$&$0.33$&$0.15\f{8}; 0.26\f{9}$&$0.37$ \\
             &$0.39\f{2}; 0.31\f{3}$ \\
$\gamma_3   $&$1.56\f{1}; 1.446\f{1}$&$1.54$&$1.55\f{8}; 1.39\f{9}$&$1.43$ \\
             &$1.44\f{2}; 1.46\f{3}$ \\

Ge & Exp. & GW & $sp^3d^5s^*$ TB & Present model\\
\hline
$m_l^{\rm L}$&$1.588\f{4}; 1.74\f{5}   $&$1.626$&$1.363\f{8}; 1.584\f{9}$&$1.594$ \\
$m_t^{\rm L}$&$0.08152\f{4}; 0.079\f{5}$&$0.074$&$0.083\f{8}; 0.081\f{9}$&$0.082$ \\
$m_l^\Delta $&$                        $&$0.881$&$0.655\f{8}; 0.701\f{9}$&$0.837$ \\
$m_t^\Delta $&$                        $&$0.176$&$0.223\f{8}; 0.201\f{9}$&$0.178$ \\
$m_\Gamma   $&$                        $&$     $&$0.038\f{8}; 0.039\f{9}$&$0.038$ \\

$\gamma_1   $&$13.0\f{6}; 12.8\f{7};   $&$13.54$&$13.13\f{8}; 13.14\f{9}$&$12.96$ \\
             &$13.25\f{1}$ \\
$\gamma_2   $&$4.4\f{6}; 4.0\f{7};     $&$4.32 $&$ 4.01\f{8};  3.68\f{9}$&$ 4.11$ \\
             &$4.20\f{1}$ \\
$\gamma_3   $&$5.3\f{6}; 5.5\f{7};     $&$5.77 $&$ 5.75\f{8};  5.63\f{9}$&$ 5.59$ \\
             &$5.56\f{1}$ \\
\end{tabular}
\end{ruledtabular}
\footnotetext[1]{Ref.~\onlinecite{Landolt}} 
\footnotetext[2]{I. Balslev and P. Lawaetz, as presented in Ref.~\onlinecite{Humphreys81}.}
\footnotetext[3]{Ref.~\onlinecite{Lawaetz71}.}
\footnotetext[4]{Ref.~\onlinecite{Levinger61}.}
\footnotetext[5]{Ref.~\onlinecite{Halpern65}.}
\footnotetext[6]{Ref.~\onlinecite{Dresselhaus55}.}
\footnotetext[7]{Ref.~\onlinecite{Makarov83}.}
\footnotetext[8]{Ref.~\onlinecite{Jancu98}.}
\footnotetext[9]{Ref.~\onlinecite{Boykin04}.}
\caption{Effective masses and Luttinger parameters of Si and Ge.}
\label{tablemasses}
\end{table}

\begin{table}
\centering
\begin{ruledtabular}
\begin{tabular}{lcccc}
Si & Exp.\f{1} & LDA\f{2} & $sp^3d^5s^*$\f{3} & Present model \\
\hline
$b_v                                       $&$-2.10\pm0.10$&$-2.27$&$-1.85$&$-2.12$ \\	
$d_v                                       $&$-4.85\pm0.15$&$-4.36$&$-5.46$&$-4.91$ \\	
$\Xi_d^\Delta+\frac{1}{3}\Xi_u^\Delta-a_v  $&$ 1.50\pm0.30$&$ 1.67$&$ 0.97$&$ 1.43$ \\
$\Xi_u^\Delta                              $&$ 8.60\pm0.40$&$ 8.79$&$ 6.88$&$ 8.70$ \\
$\Xi_d^{\rm L}+\frac{1}{3}\Xi_u^{\rm L}-a_v$&$            $&$-3.14$&$-2.61$&$-3.20$ \\
$\Xi_u^{\rm L}                             $&$            $&$13.85$&$ 3.69$&$16.19$ \\

Ge & Exp.\f{1} & LDA\f{2} & $sp^3d^5s^*$\f{3} & Present model \\
\hline
$b_v                                       $&$-2.86\pm0.15$&$-2.90$&$-2.48$&$-2.74$ \\
$d_v                                       $&$-5.28\pm0.50$&$-6   $&$-3.74$&$-5.09$ \\
$a_g(\Gamma)                               $&$            $&$     $&$-9.54$&$-9.01$ \\
$\Xi_d^{\rm L}+\frac{1}{3}\Xi_u^{\rm L}-a_v$&$-2.00\pm0.50$&$-2.86$&$-2.85$&$-3.19$ \\
$\Xi_u^{\rm L}                             $&$16.20\pm0.40$&$17   $&$ 8.09$&$15.39$ \\
$\Xi_d^\Delta+\frac{1}{3}\Xi_u^\Delta-a_v  $&$            $&$ 1.43$&$ 3.50$&$ 1.10$ \\
$\Xi_u^\Delta                              $&$            $&$10   $&$ 6.50$&$ 9.02$ \\
\end{tabular}
\end{ruledtabular}
\footnotetext[1]{Cited by Ref.~\onlinecite{Walle86}.}
\footnotetext[2]{Present work.}
\footnotetext[3]{Refs.~\onlinecite{Boykin07} and \onlinecite{Boykin04}.}
\caption{Deformation potentials of Si and Ge (eV).}
\label{tablevdef}
\end{table}

The TB deformation potentials of the conduction and valence band extrema of Si and Ge are listed in Table \ref{tablevdef}, and compared with the experimental and LDA data. The TB model performs well on all relevant deformation potentials. Also shown are the results obtained with the $sp^3d^5s^*$ model and parameterization of Refs. \onlinecite{Boykin07} and \onlinecite{Boykin04}. The present model reproduces the uniaxial $\langle111\rangle$ deformation potentials $d_v$ and $\Xi_u^{\rm L}$ significantly better, as it is able to account for the on-site couplings between the orbitals under shear strains. The hydrostatic valence band deformation potential $a_v$, which controls the position of the band structure on an {\it absolute} energy scale, has been fitted to Ref. \onlinecite{Li06} ($a_v=2.38$ eV for Si and $a_v=2.23$ eV for Ge). Accordingly, the {\it unstrained} valence band offset between Si and Ge has been set to $\Delta_{\rm VBO}=0.68$ eV, to reproduce the experimental valence band discontinuity between Si$_{1-x}$Ge$_x$ alloys and Si as best as possible (see Table \ref{tableparamsSiGe} and paragraph \ref{subsectionalloys}). We achieve that way {\it strained} valence band offsets $\Delta_{\rm VBO}=0.79$ eV on Si $[001]$ and $\Delta_{\rm VBO}=0.28$ eV on Ge $[001]$, within the experimental error bars\cite{Schwartz89} and in-between the theoretical LDA values of Refs. \onlinecite{Walle86} and \onlinecite{Colombo91}. The unstrained valence band offset and hydrostatic deformation potential, which are still somewhat controversial,\cite{Walle86,Cardona87,Walle89,Resta90,Wei99,Li06} can be tuned by shifting all on-site energies and $\alpha$'s.

\begin{figure}
\centering
\includegraphics[width=0.50\columnwidth]{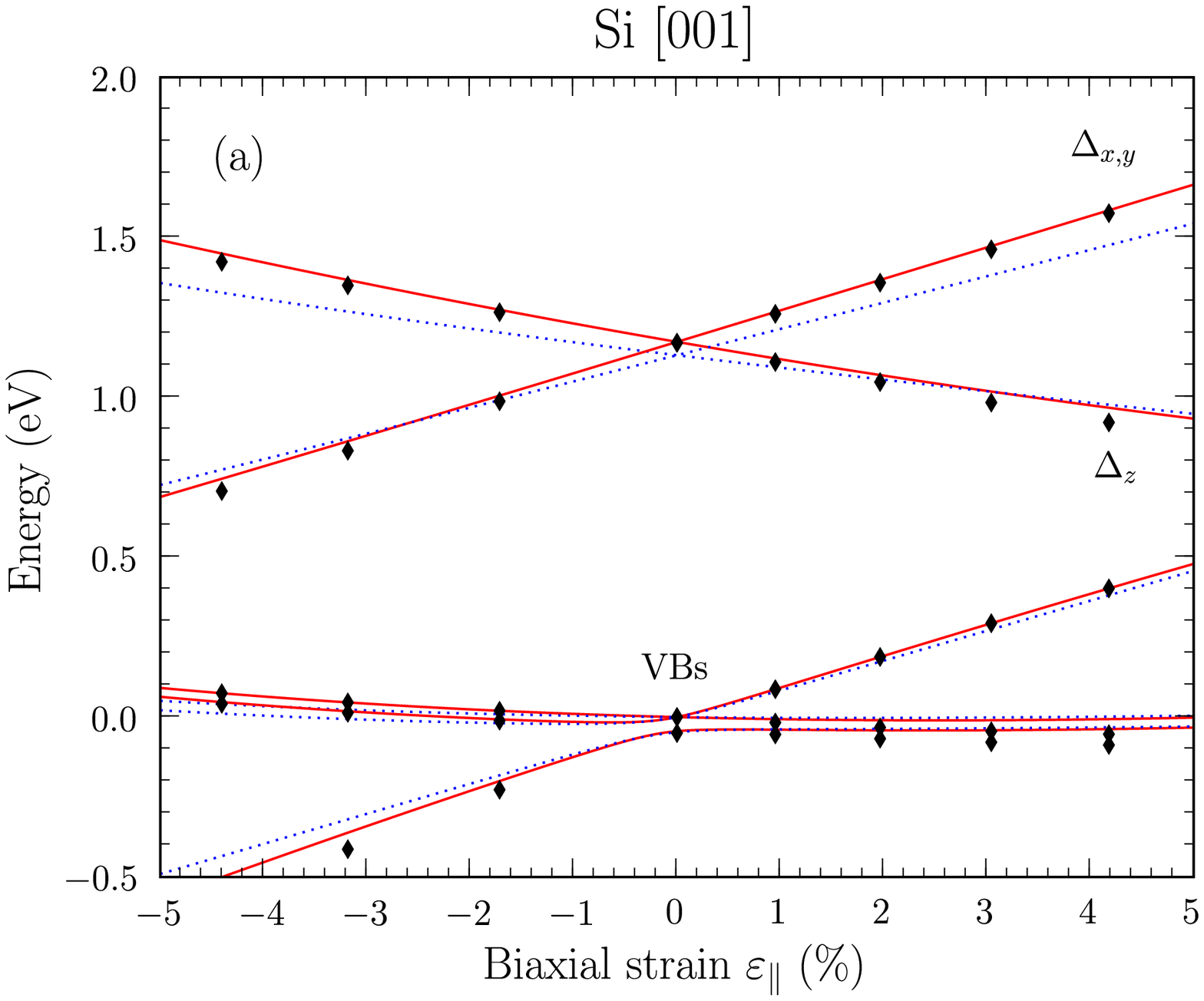}\\
\includegraphics[width=0.50\columnwidth]{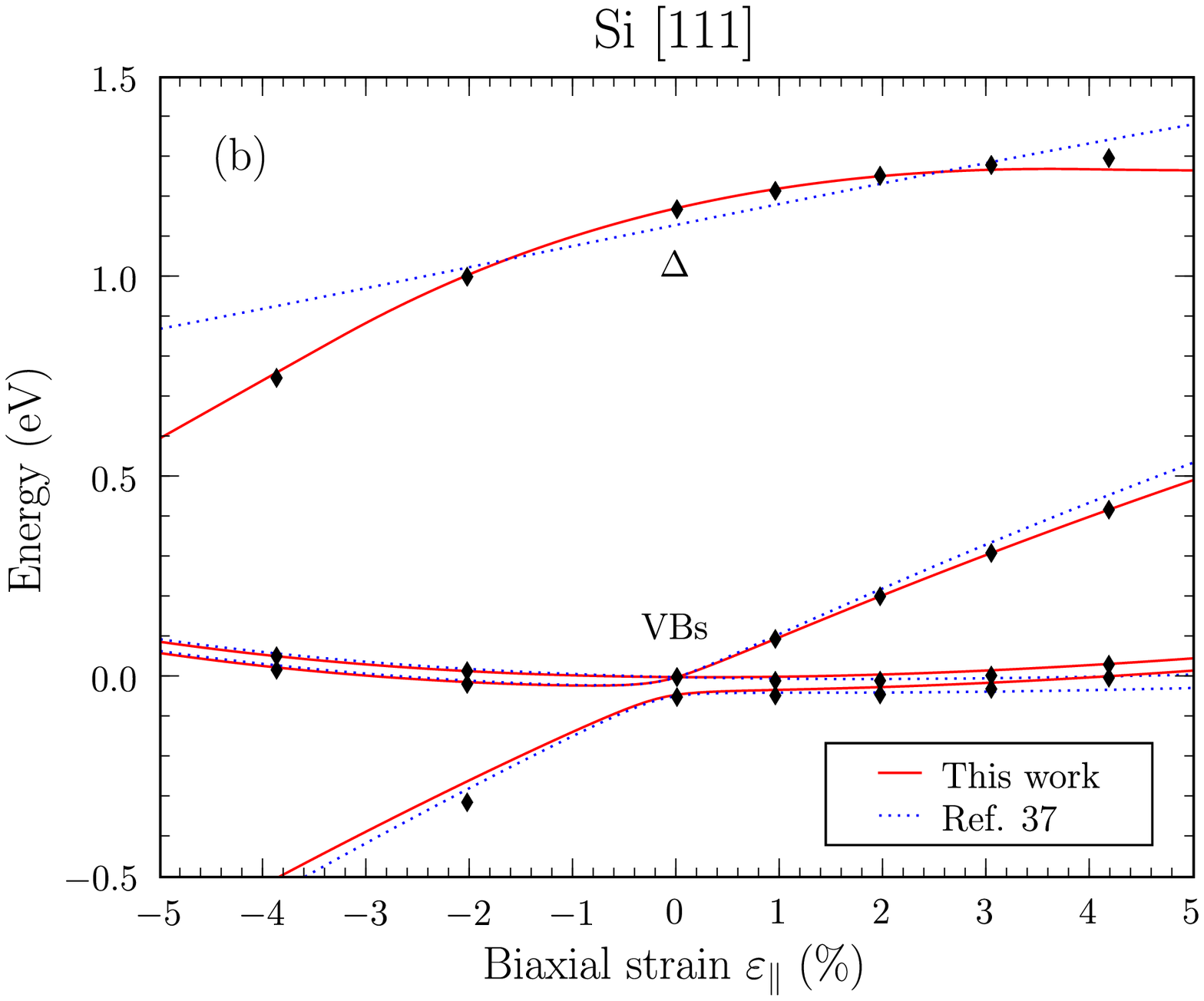}\\
\includegraphics[width=0.50\columnwidth]{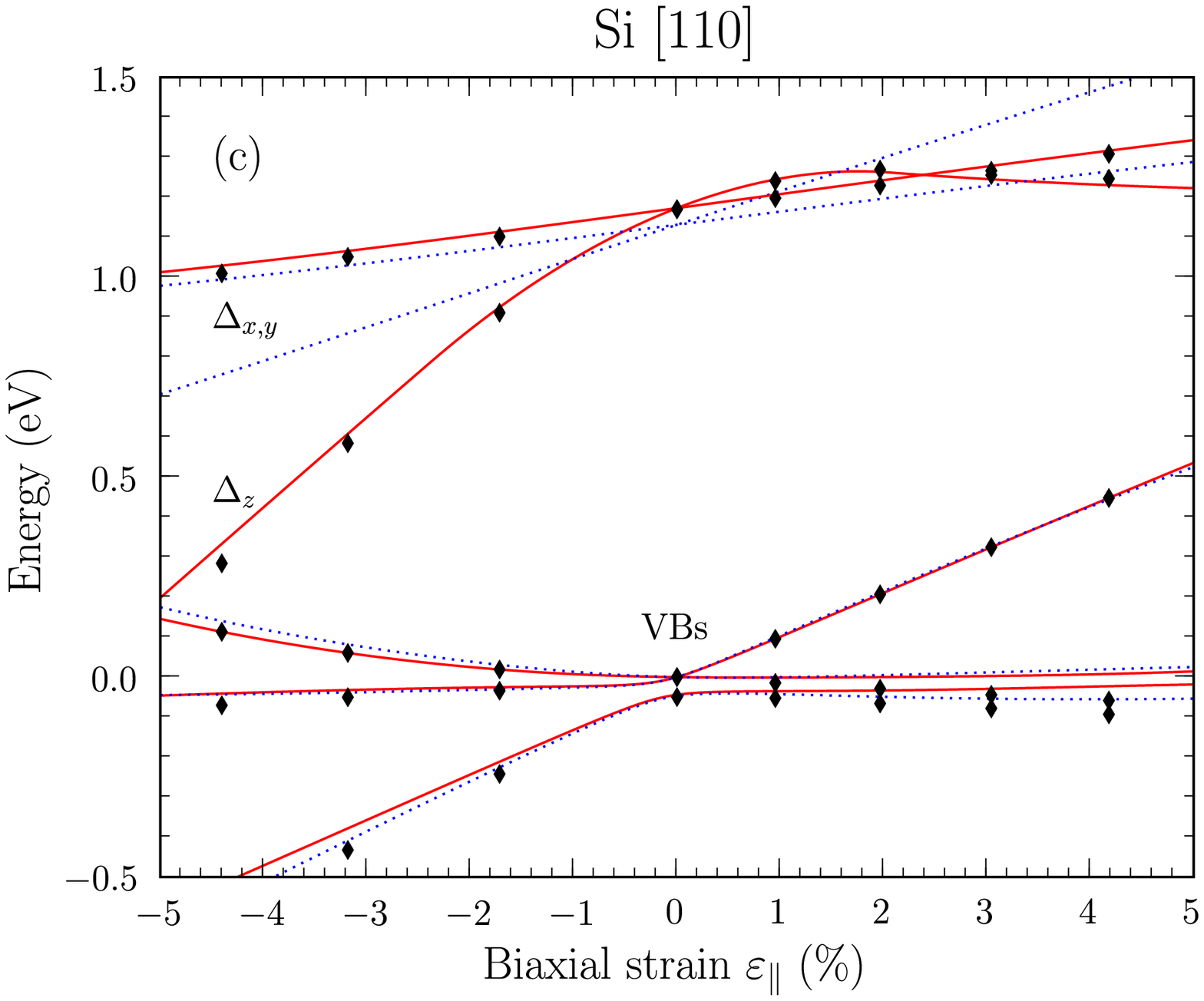}
\caption{(Color online) (a) $[001]$, (b) $[111]$ and (c) $[110]$ biaxial strain behavior of bulk silicon.\cite{noteuniax} The black diamonds are the {\it ab initio} data.\cite{noteavlda}}
\label{figUniaxSi}
\end{figure}

\begin{figure}
\centering
\includegraphics[width=0.50\columnwidth]{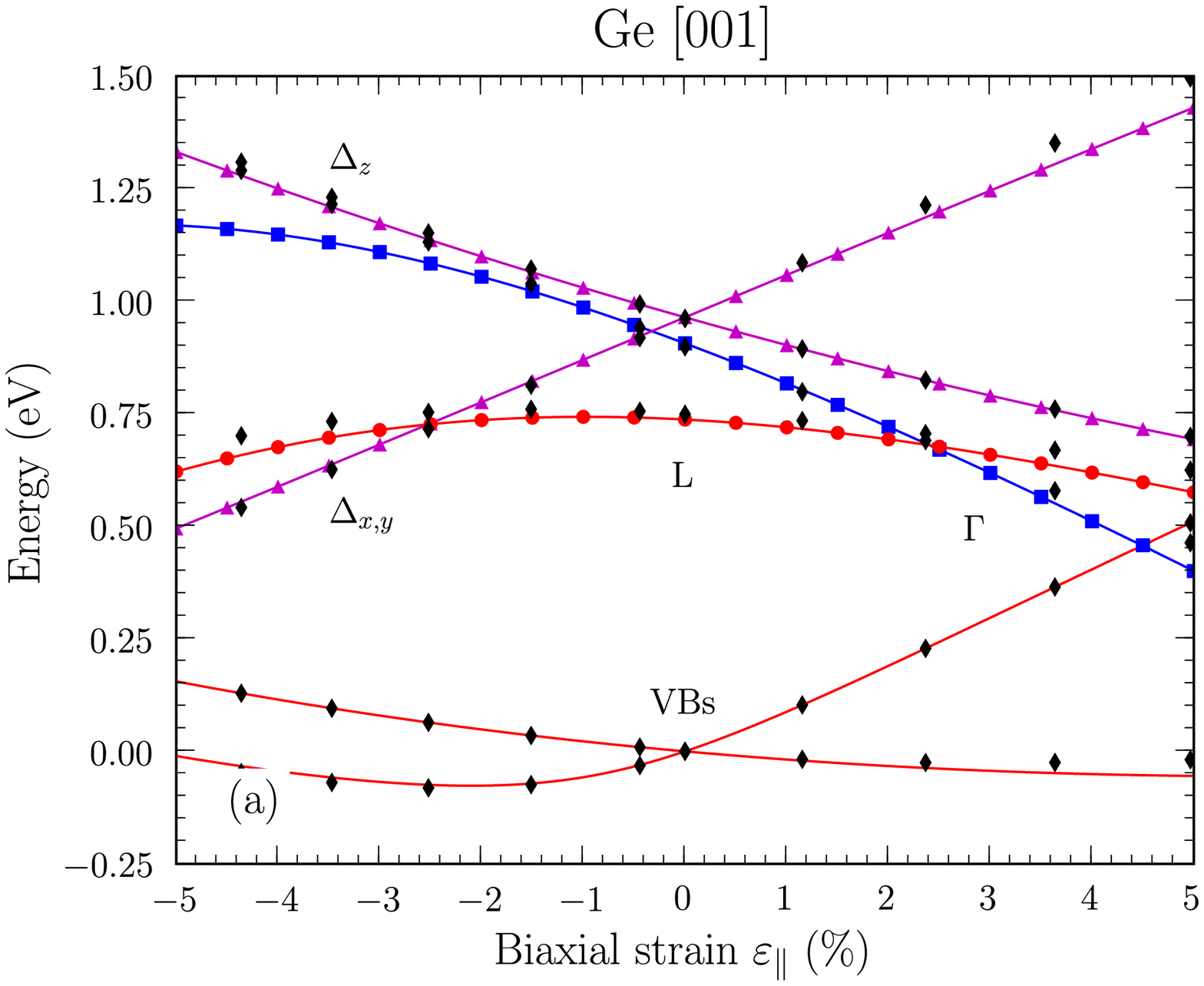}\\
\includegraphics[width=0.50\columnwidth]{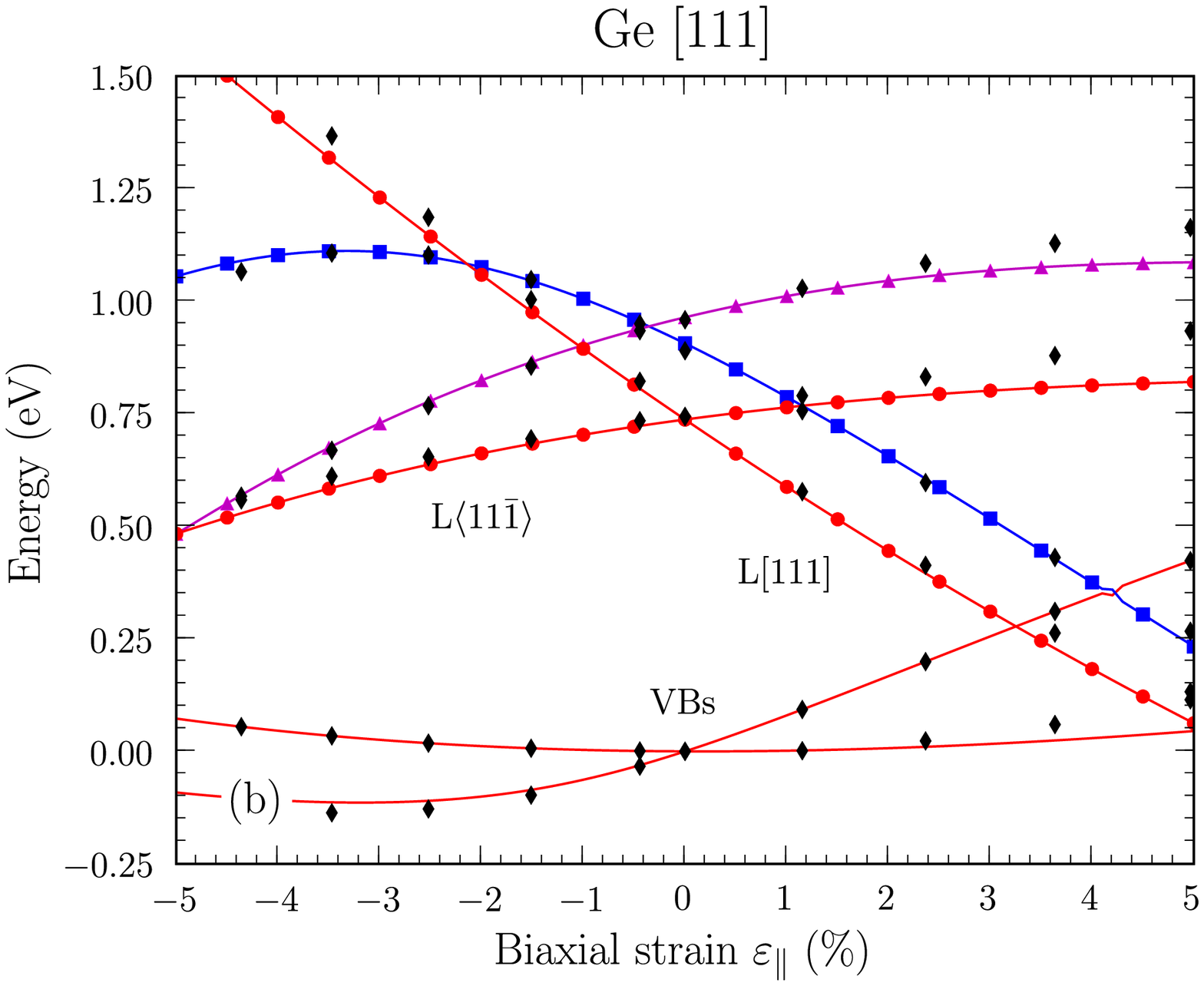}\\
\includegraphics[width=0.50\columnwidth]{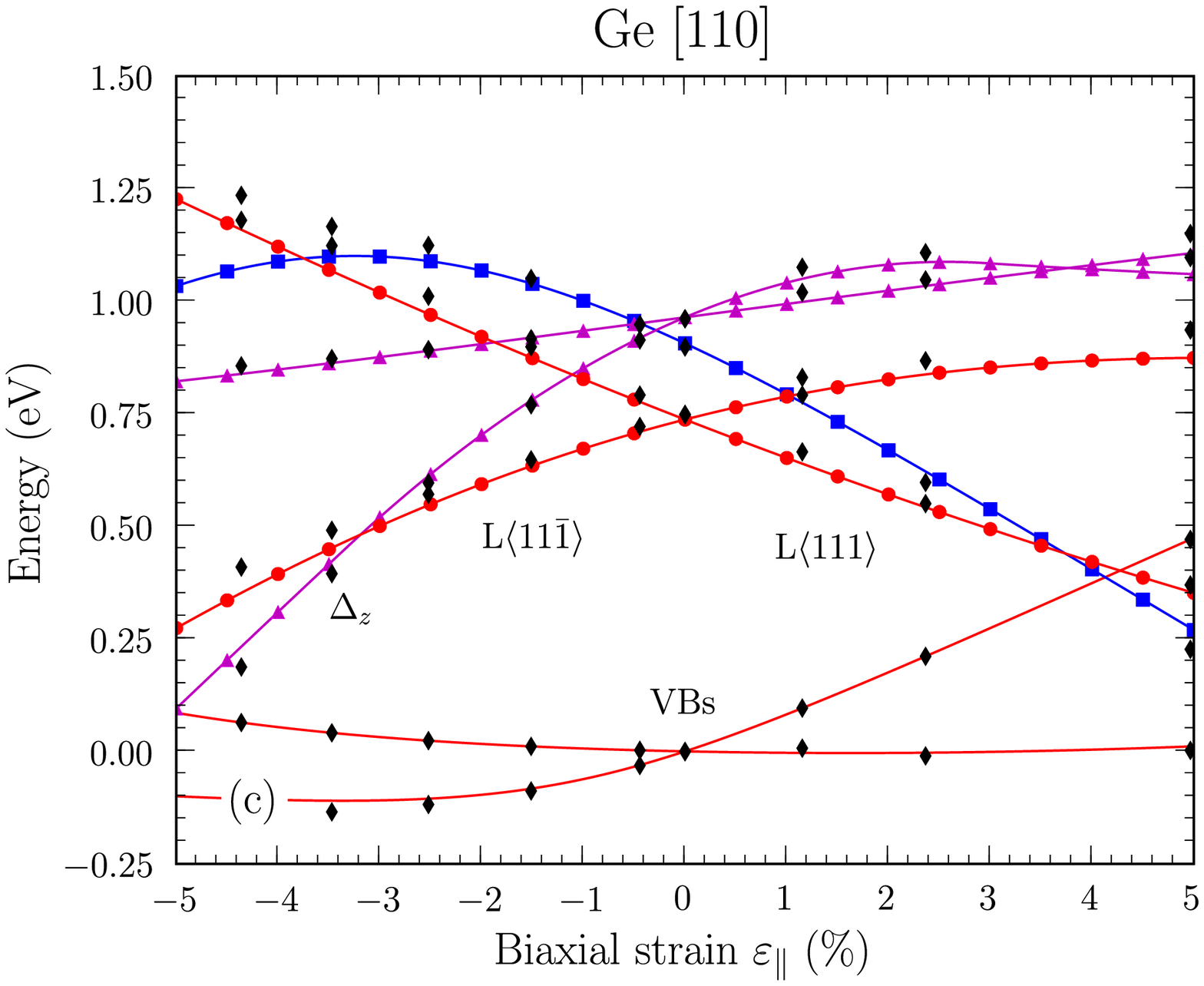}
\caption{(Color online) (a) $[001]$, (b) $[111]$ and (c) $[110]$ biaxial strain behavior of bulk germanium.\cite{noteuniax} The (red) dots, (blue) squares and (magenta) triangles are the L, $\Gamma$ and $\Delta$ valleys respectively. The black diamonds are the {\it ab initio} data.\cite{noteavlda}}
\label{figUniaxGe}
\end{figure}

The typical behavior of the valence and conduction bands of Si and Ge under biaxial $[001]$, $[110]$ and $[111]$ strain is plotted as a function of the in-plane deformation $\varepsilon_\parallel$ in Figs. \ref{figUniaxSi} and \ref{figUniaxGe}.\cite{noteuniax} As a reference, the lattice mismatch of Si grown on Ge is $\varepsilon_\parallel=4.18$ \%, while the lattice mismatch of Ge grown on Si is $\varepsilon_\parallel=-4.01$\%. Under biaxial $[001]$ strain, the six conduction band minima of silicon split in two groups,\cite{Walle89} the $\Delta_{x,y}$ valleys oriented along $[100]$ and $[010]$, and the $\Delta_z$ valleys oriented along $[001]$. The conduction band edges are almost linear with strain, $\Delta_{x,y}$ being the lowest energy valleys when $\varepsilon_\parallel<0$, and $\Delta_{z}$ when $\varepsilon_\parallel>0$. Biaxial $[110]$ strain also splits the conduction band minima the same way; the $\Delta_z$ minima are, however, markedly non-linear, being the lowest energy valleys for both $\varepsilon_\parallel<0$ and $\varepsilon_\parallel\gtrsim2$ \%. This behavior, which is not accounted for by the simplest deformation potential theories, results from the shear strains component $\varepsilon_{xy}=(\varepsilon_\perp-\varepsilon_\parallel)/2$ (see discussion below). As a matter of fact, biaxial $[111]$ strain (which does not split the conduction band minima\cite{Walle89}), also exhibits the same non-linear trends [$\varepsilon_{yz}=\varepsilon_{xz}=\varepsilon_{xy}=(\varepsilon_\perp-\varepsilon_\parallel)/3$]. It is worthwile to note that these non-linearities have not been specifically targeted in the optimization of the TB parameters.

In Germanium, a biaxial $[001]$ strain likewise splits the $\Delta$ valleys (but not the L ones). The $\Delta_{x,y}$ valleys are the lowest energy bands for compressive strain $\varepsilon_\parallel\lesssim2$ \%, while the $\Gamma$ valley falls below the L valleys for tensile strain $\varepsilon_\parallel\gtrsim2$ \%. Germanium then becomes a small, direct band gap semiconductor, and even a semi-metal (zero gap) when $\varepsilon_\parallel\gtrsim4$ \%. Biaxial $[111]$ strain splits the L valleys in two groups,\cite{Walle89} the three L$\langle11\bar1\rangle$ valleys (lowest energy for compressive strains) and the L$[111]$ valley (lowest energy for tensile strains). The band gap again closes when $\varepsilon_\parallel\gtrsim3.5$ \%. The behavior of germanium under biaxial $[110]$ strain is much more complex, $\Delta_{z}$, the two L$\langle11\bar1\rangle$ valleys, the two L$\langle111\rangle$ valleys and $\Gamma$ being successively the lowest energy band(s) when going from compressive to tensile strains, with a zero band gap for $\varepsilon_\parallel\lesssim-5$ \% and $\varepsilon_\parallel\gtrsim4$ \%.

\begin{figure}
\centering
\includegraphics[width=0.66\columnwidth]{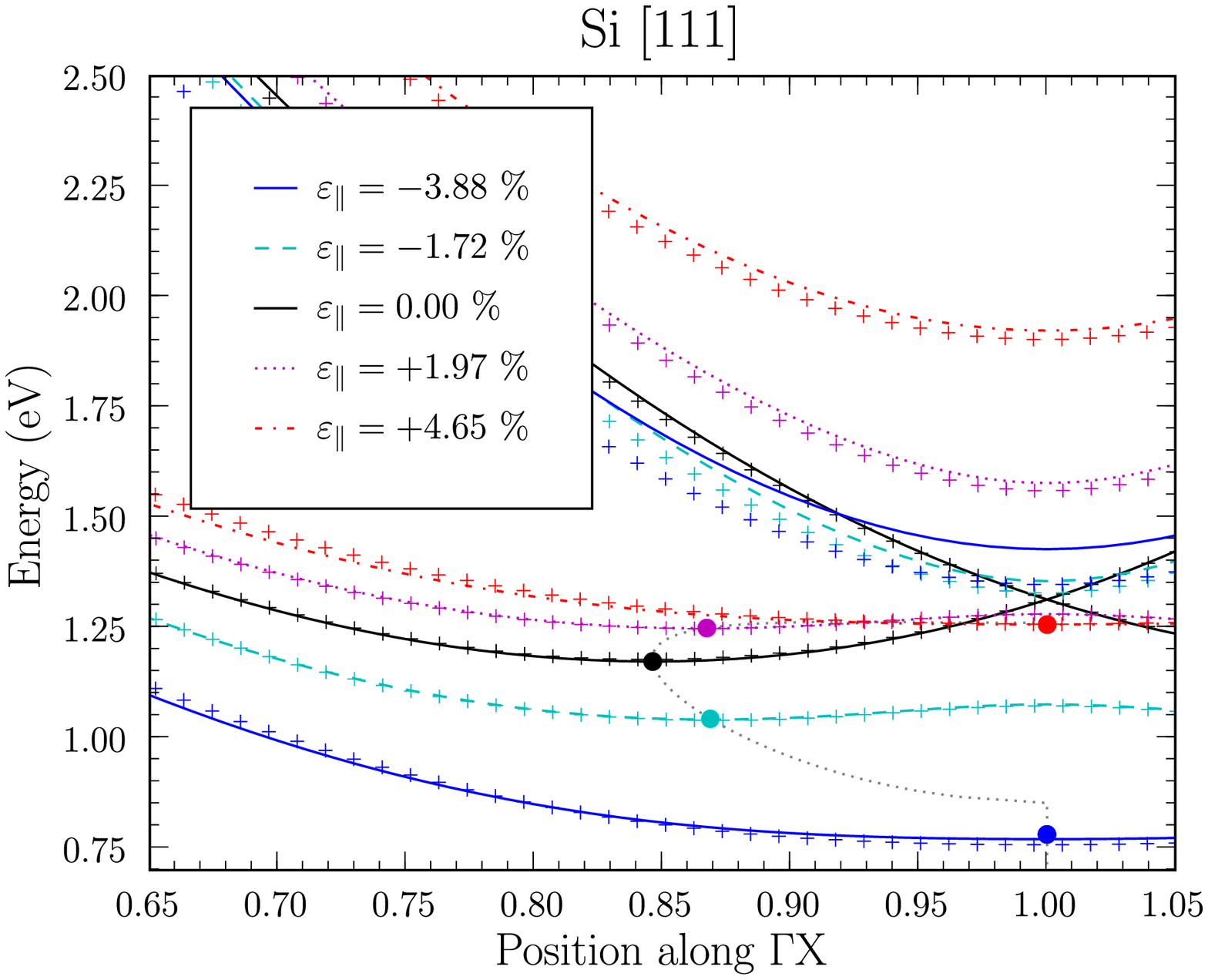}
\caption{(Color online) Lowest two conduction bands (plotted along the $\Gamma$X axis) of bulk silicon under $[111]$ biaxial strain. The conduction band minimum is marked with a dot. Its motion with strain is plotted as a dotted gray line. The crosses are the {\it ab initio} data (also see Fig. \ref{figMassesSi}).}
\label{figBCSi}
\end{figure}

\begin{figure}
\centering
\includegraphics[width=0.66\columnwidth]{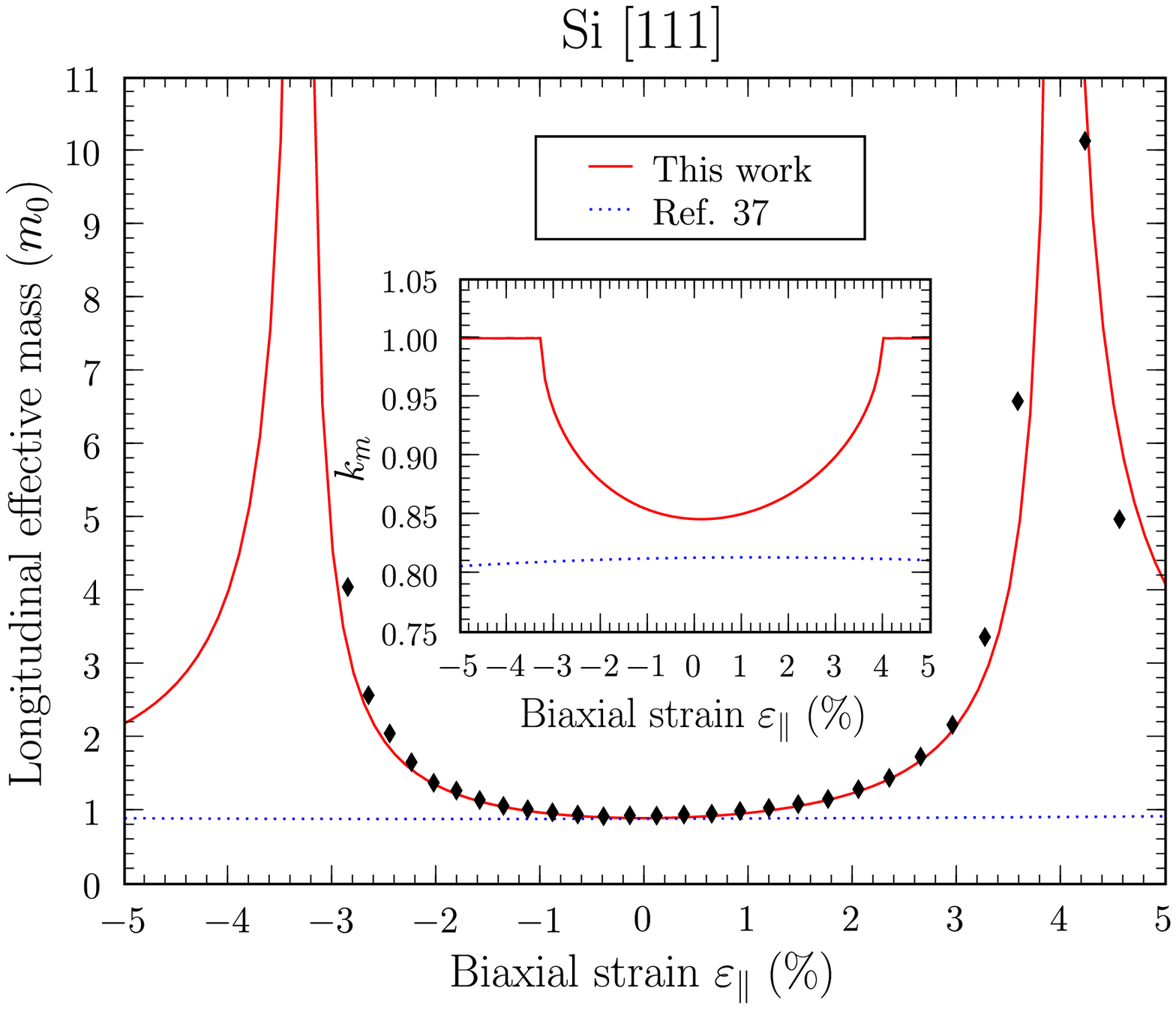}
\caption{(Color online) Conduction band, longitudinal effective mass of bulk silicon under $[111]$ biaxial strain. Inset: Position of the conduction band minimum along the $\Gamma$X axis. The black diamonds are the {\it ab initio} data.}
\label{figMassesSi}
\end{figure}

These results are in good agreement with the {\it ab initio} data (black diamonds\cite{noteavlda} on Figs. \ref{figUniaxSi} and \ref{figUniaxGe}). Though the $sp^3d^5s$ parametrization of Refs. \onlinecite{Boykin07} and \onlinecite{Boykin04} also shows reasonnable agreement with {\it ab initio} data for $[001]$ biaxial strain, it notably misses the strong non-linearity of the conduction band energy in Si $[110]$ or $[111]$. As stated previously, this non-linearity results from the peculiar behavior of the $\Delta$ valleys under shear strains, as further evidenced in Fig. \ref{figBCSi}. Indeed, the conduction band minima move towards the X points with increasing compressive or tensile $[111]$ strain, and finally hang to the latter when $\varepsilon_\parallel>4$ \% or $\varepsilon_\parallel<-3.3$ \%.\cite{Rideau06} The position of the conduction band minima along the $\Gamma$X axis, as well as the longitudinal effective mass are plotted in Fig. \ref{figMassesSi}. The longitudinal effective mass dramatically increases with $|\varepsilon_\parallel|$, and eventually diverges (quartic conduction band dispersion) before decreasing again when the conduction band minima reach the edge of the first Brillouin zone. Likewise, the $\Delta_z$ valleys shift to the X point under biaxial $[110]$ strain, with a divergence of the longitudinal effective mass at $\varepsilon_\parallel\simeq2.1$ \% and $\varepsilon_\parallel\simeq-2.4$ \%.\cite{Rideau06,Ungersboeck07,Yamauchi08} The splitting of the transverse masses under uniaxial $\langle110\rangle$ strains is also well reproduced.\cite{Rideau06} These effects, which were not accounted for by previous TB parametrizations, are fundamental for the understanding of the transport properties of strained MOSFETS or SiGe nanowire heterostructures.\cite{Thompson04,Irie04,Ungersboeck07}

\subsection{Results in disordered Si$_{1-x}$Ge$_{x}$ alloys}
\label{subsectionalloys}

\begin{table}
\centering
%\newcolumntype{d}[1]{D{.}{.}{#1}}
\begin{ruledtabular}
\begin{tabular}{cdddddd}
Material  & \multicolumn{1}{c}{$\alpha$ (N/m)} & \multicolumn{1}{c}{$\beta$ (N/m)} & \multicolumn{1}{c}{$c_{11}$ (GPa)} & \multicolumn{1}{c}{$c_{12}$ (GPa)} & \multicolumn{1}{c}{$c_{44}$ (GPa)} & \multicolumn{1}{c}{$\zeta$} \\
\hline
Si & 48.54 & 13.83 & 165.8 & 63.9 & 79.3 & 0.557 \\
Ge & 39.14 & 11.81 & 131.8 & 48.3 & 64.1 & 0.536 \\
Si$_{0.5}$Ge$_{0.5}$ & 43.80 & 12.81 & 148.5 & 55.9 & 71.6 & 0.547 \\
\end{tabular}
\end{ruledtabular}
\caption{The valence force field bond-stretching constant $\alpha$, bond-bending constant $\beta$, elastic constants $c_{ij}$ and internal strain parameter of Si, Ge and Si$_{0.5}$Ge$_{0.5}$. In the disordered SiGe alloys, we choose $\beta=13.31$ N/m for Si$-$Si$-$Ge and $\beta=12.30$ N/m for Ge$-$Ge$-$Si pairs of bonds.}
\label{tablekeating}
\end{table}

The TB method offers the opportunity to describe semiconductor alloys as random distributions of atoms instead of virtual crystals, thus allowing, e.g., the investigation of alloy disorder scattering. The present TB model is particularly well suited to the such random alloys as it does not depend on macroscopic strains that would be ill-defined in a disordered environment. We have therefore computed the band gap energy of bulk Si$_{1-x}$Ge$_{x}$ alloys modeled as random distributions of Si and Ge atoms in large $\sim 65000$ atoms supercells (in order to reduce the statistical noise). The lattice parameters of these supercells and the internal coordinates of the atoms were optimized with Keating's valence force field model.\cite{Keating66} The bond bending and bond strecthing constants of the SiGe alloy are given in Table \ref{tablekeating}.

\begin{figure}
\centering
\includegraphics[width=0.66\columnwidth]{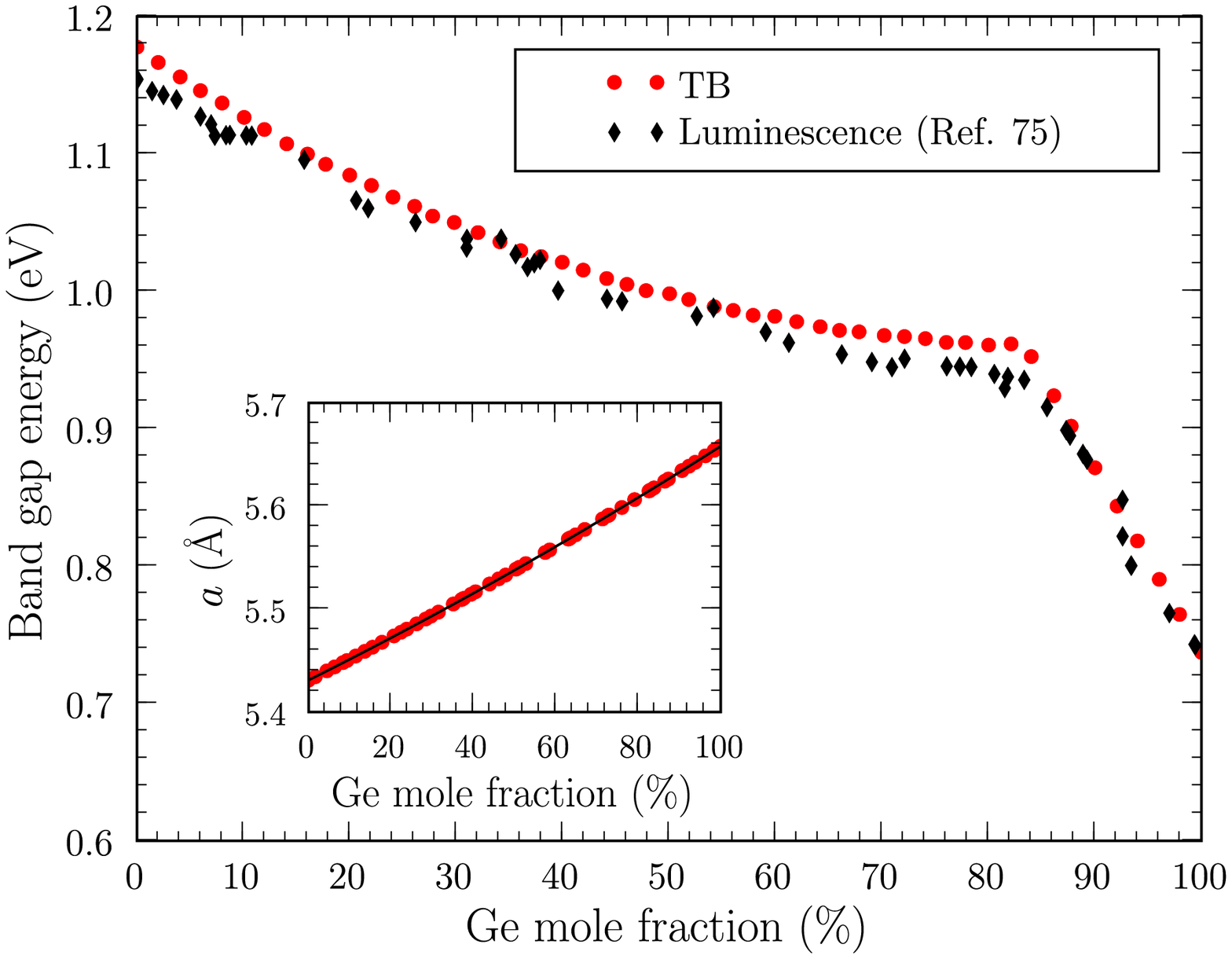}
\caption{(Color online) Band gap energy of random, bulk Si$_{1-x}$Ge$_x$ alloys as a function of the Ge mole fraction $x$. Inset: Lattice parameter of the alloy. The solid line is Dismukes's law\cite{Dismukes64} $a(x)=5.431+0.2x+0.027x^2$ \AA.}
\label{figgapSiGe}
\end{figure}

The calculated band gap energy of the alloy is plotted as a function of the Ge mole fraction $x$ in Fig. \ref{figgapSiGe}, and compared with luminescence data.\cite{Weber89} The lattice parameter of the alloy (computed from the valence force field) is also plotted in the inset, and matches Dismukes's law\cite{Dismukes64} $a(x)=5.431+0.2x+0.027x^2$ \AA\ (solid line). The present model predicts a crossing between the $\Delta$ and L-valley conduction band minima around $x=0.84$, in agreement with the experimental data. The bowing of the band gap energy for $x<0.84$ is, in particular, very well reproduced by the tight-binding calculation. We find that the band gap energy of the disordered Si$_{0.5}$Ge$_{0.5}$ alloy is only $\simeq 5$ meV lower than the band gap energy of the ordered alloy.

\begin{figure}
\centering
\includegraphics[width=0.66\columnwidth]{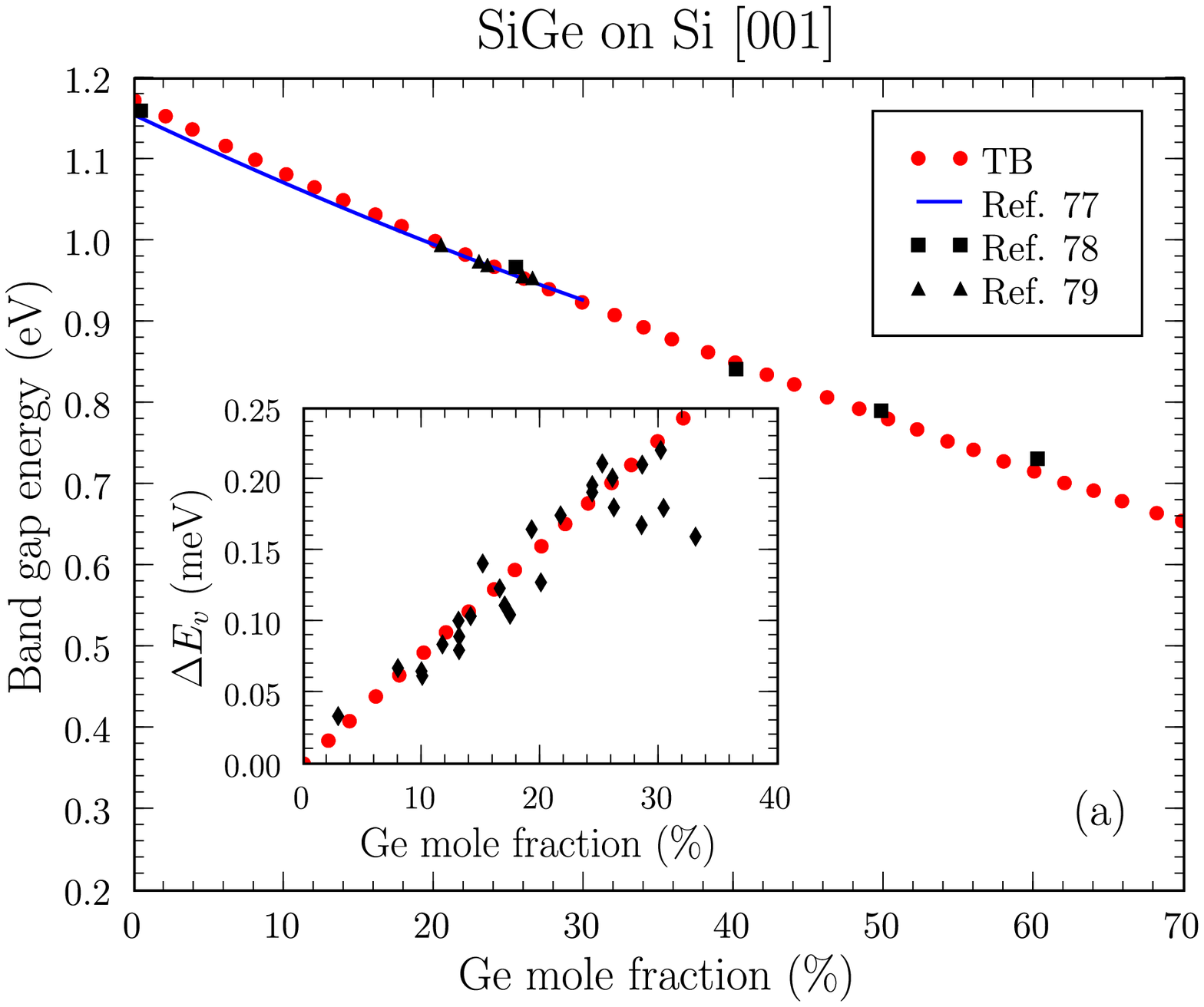} \\
\includegraphics[width=0.66\columnwidth]{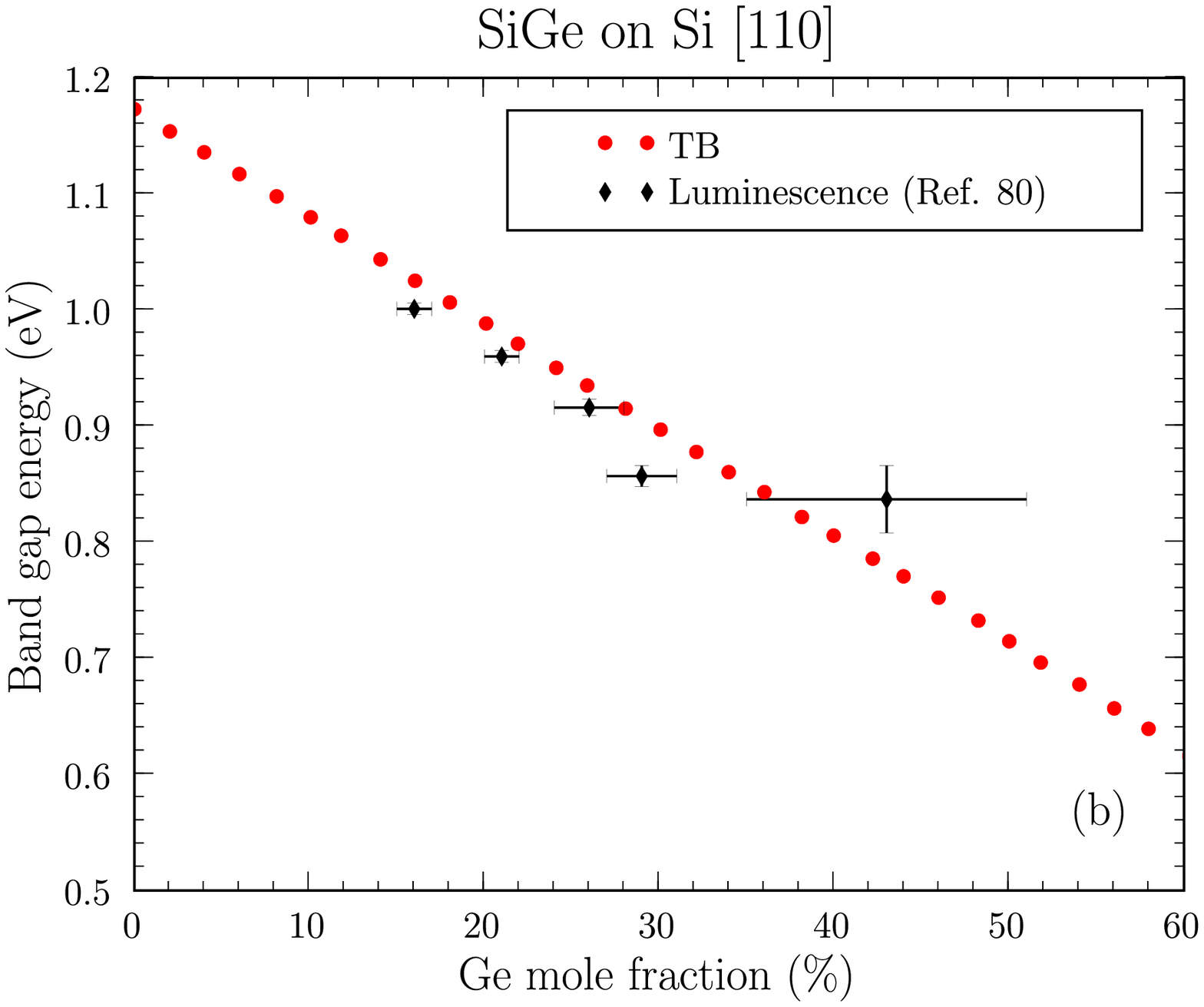}
\caption{(Color online) (a) Band gap energy of random Si$_{1-x}$Ge$_x$ alloys biaxially strained on Si [001], as a function of the Ge mole fraction $x$. The inset is the TB valence band discontinuity $\Delta E_v$ (dots) compared to various experimental sources (diamonds) compiled in Ref. \onlinecite{Takagi98}. (b) Band gap energy of random Si$_{1-x}$Ge$_x$ alloys biaxially strained on Si [110].}
\label{figgapstrainedSiGe}
\end{figure}

As another illustration, the band gap energy of random SiGe alloys biaxially strained on Si [001] or [110] is plotted in Fig. \ref{figgapstrainedSiGe}. Also shown in the inset of Fig. \ref{figgapstrainedSiGe}(a) is the valence band discontinuity in Si$_{1-x}$Ge$_x$ alloys on Si [001]. The band gap decreases much faster with the Ge mole fraction than in bulk alloys due to the strains. This decrease is again very well reproduced by the TB model, showing its ability to describe random alloys under arbitrary strains. The conduction and valence band offsets of disordered Si$_{1-x}$Ge$_x$ alloys on Si$_{1-y}$Ge$_y$ [001] buffers are likewise in agreement with the 30 bands $\vec{k}\cdot\vec{p}$ model of Ref. \onlinecite{Rideau06} in the virtual crystal approximation.

\section{Conclusion}

We have presented a model for the on-site matrix elements of the $sp^3d^5s^*$ TB hamiltonian of a strained diamond or zinc-blende crystal. This model improves over previous parametrizations by including the on-site couplings between the $s$, $p$ and between the $d$ orbitals of the atoms under uniaxial and shear strains. It is able to reproduce the deformation potentials and the dependence of the effective masses on strains at all relevant k-points of the first Brillouin zone and is fully consistent with the symmetries of the crystal. We have succesfully applied this description to Si, Ge and their alloys. This tight-binding model should allow predictive modeling of the electronic properties of strained Si/Ge heterostructures, and is numerically efficient enough to be included, e.g., in full-band Monte-Carlo\cite{Fischetti88} or Kubo-Greenwood\cite{Pham07} calculations of the transport properties of semiconductor devices.

\section*{Acknowledgements}

This work was supported by the french national research agency (ANR) project ``QuantaMonde'' (contract ANR-07-NANO-023-02). We thank C. Delerue and J.-M. Jancu for reading the manuscript and for fruitful discussions about tight-binding.

\appendix

\section{Complete on-site couplings between the $s$ ($s^*$), $p$, and $d$ orbitals}
\label{AppendixCouplings}

\subsection{On-site energy of the $s$ ($s^*$) orbitals}

The on-site energy of the $s$ orbitals reads:
\begin{equation}
H_s=E_s^0+\alpha_s\frac{\Delta\Omega}{\Omega_0},
\label{eqsorbs}
\end{equation}
where $E_s^0$ is the unstrained $s$ orbital energy and $\alpha_s$ characterizes the dependence of $H_s$ on the hydrostatic strain. The same model applies to the $s^*$ orbitals (possibly with a different $\alpha_s^*$ coefficient).

\subsection{On-site couplings between the $d$ orbitals}
\label{subsectiond}

The on-site $d$ block matrix $\hat{\vec{H}}_d$ reads in the $\{d_{yz},d_{xz},d_{xy},d_{x^2-y^2},d_{3z^2-r^2}\}$ basis set:
\begin{widetext}
\begin{eqnarray}
\hat{\vec{H}}_d&=&\left(E_d^0+\alpha_d\frac{\Delta\Omega}{\Omega_0}\right)\hat{\vec{I}}+\sum_j^{\rm NN}\beta_d(d)
\begin{bmatrix}
l^2-\frac{1}{3} & -lm & -ln & mn & -\frac{1}{\sqrt{3}}mn \\
-lm & m^2-\frac{1}{3} & -mn & -ln & -\frac{1}{\sqrt{3}}ln \\
-ln & -mn & n^2-\frac{1}{3} & 0 & \frac{2}{\sqrt{3}}lm \\
mn & -ln & 0 & n^2-\frac{1}{3} & \frac{2}{\sqrt{3}}u \\
-\frac{1}{\sqrt{3}}mn & -\frac{1}{\sqrt{3}}ln & \frac{2}{\sqrt{3}}lm & \frac{2}{\sqrt{3}}u & -n^2+\frac{1}{3} \\
\end{bmatrix} \\
&+&\sum_j^{\rm NN}\gamma_d(d)
\begin{bmatrix}
m^2n^2 & lmn^2 & lm^2n & mnu & mnv \\
lmn^2 & l^2n^2 & l^2mn & lnu & lnv \\
lm^2n & l^2mn & l^2m^2 & lmu & lmv \\
mnu & lnu & lmu & u^2 & uv \\
mnv & lnv & lmv & uv & v^2 \nonumber \\
\end{bmatrix},
\label{eqdorbs}
\end{eqnarray}
\end{widetext}
where $E_d^0$ is the ``bare'' $d$ orbital energy (see discussion below), $u=(l^2-m^2)/2$ and $v=(3n^2-1)/(2\sqrt{3})$. Like $\beta_p(d)$, $\beta_d(d)$ and $\gamma_d(d)$ can be written:
\begin{subequations}
\begin{eqnarray}
\beta_d(d)&=&\left\langle d_i^\delta\left|\nu_2\right|d_i^\delta\right\rangle-\left\langle d_i^\pi\left|\nu_2\right|d_i^\pi\right\rangle=\beta_d^{(0)}+\beta_d^{(1)}\frac{d-d_0}{d_0} \\
\gamma_d(d)&=&3\left\langle d_i^\sigma\left|\nu_2\right|d_i^\sigma\right\rangle+\left\langle d_i^\delta\left|\nu_2\right|d_i^\delta\right\rangle-4\left\langle d_i^\pi\left|\nu_2\right|d_i^\pi\right\rangle=\gamma_d^{(0)}+\gamma_d^{(1)}\frac{d-d_0}{d_0},
\end{eqnarray}
\end{subequations}
where $\beta_d^{(0)}$, $\beta_d^{(1)}$, $\gamma_d^{(0)}$ and $\gamma_d^{(1)}$ are additional TB parameters.

$\hat{\vec{H}}_d$ is the sum of the bare $d$ orbital energies, of a $\propto\alpha_d$ hydrostatic correction, and of two $\propto\beta_d,\gamma_d$ angular matrices. Like $\hat{\vec{H}}_p$, the diagonal of the $\propto\beta_d$ matrix has been shifted [by $-1/3\sum_{NN}\beta_d(d)$] so as to be zero in the unstrained diamond or zinc-blende crystal. The five $d$ orbitals are not, however, degenerate any more as soon as $\gamma_d^{(0)}\neq 0$. They indeed split in two groups: {\it i}) the $\{d_{yz},d_{xz},d_{xy}\}$ orbitals with energy $E_d^0+4\gamma_d^{(0)}/9$, and {\it ii}) the $\{d_{x^2-y^2},d_{3z^2-r^2}\}$ orbitals with energies $E_d^0$ (since, e.g., $\sum_j^{\rm NN}m^2n^2=4/9$).\cite{notecfp} This is consistent with the symmetry of the zinc-blende or diamond lattice, though it is usually not accounted for in TB models.

\subsection{Couplings between the $s$ and $s^*$ orbitals}

The on-site matrix element coupling the $s$ and $s^*$ orbitals reads:
\begin{equation}
H_{ss^*}=\sum_j^{\rm NN}\gamma_{ss^*}(d),
\label{eqstorbs}
\end{equation}
where $\gamma_{ss^*}(d)=\left\langle s_i\left|\nu_2\right|s_i^*\right\rangle=\gamma_{ss^*}^{(0)}+\gamma_{ss^*}^{(1)}\frac{d-d_0}{d_0}$. It is non zero in the unstrained diamond or zinc-blende crystal if $\gamma_{ss^*}^{(0)}=\ne0$.

\subsection{Couplings between the $s$ ($s^*$) and $p$ or $d$ orbitals}

The on-site matrices coupling the $s$ and $p$/$d$ orbitals read:
\begin{equation}
\hat{\vec{H}}_{sp}=\sum_j^{\rm NN}\beta_{sp}(d)
\begin{bmatrix}
l & m & n
\end{bmatrix}
\end{equation}
\begin{equation}
\hat{\vec{H}}_{sd}=\sum_j^{\rm NN}\beta_{sd}(d)
\begin{bmatrix}
mn & ln & lm & u & v
\end{bmatrix},
\end{equation}
where $\beta_{sp}(d)=\left\langle s_i\left|\nu_2\right|p_i^\sigma\right\rangle=\beta_{sp}^{(0)}+\beta_{sp}^{(1)}\frac{d-d_0}{d_0}$ and
$\beta_{sd}(d)=\sqrt{3}\left\langle s_i\left|\nu_2\right|d_i^\sigma\right\rangle=\beta_{sd}^{(0)}+\beta_{sd}^{(1)}\frac{d-d_0}{d_0}$. Both matrices are zero in the unstrained diamond or zinc-blende crystal.

\subsection{Couplings between the $p$ and $d$ orbitals}

The on-site matrix coupling the $p$ and $d$ orbitals reads:
\begin{eqnarray}
\hat{\vec{H}}_{pd}&=&\sum_j^{\rm NN}\beta_{pd}(d)
\begin{bmatrix}
0 & n & m &  l & -\frac{1}{\sqrt{3}}l \\
n & 0 & l & -m & -\frac{1}{\sqrt{3}}m \\
m & l & 0 &  0 &  \frac{2}{\sqrt{3}}n \\
\end{bmatrix} \nonumber \\
&+&\sum_j^{\rm NN}\gamma_{pd}(d)
\begin{bmatrix}
lmn  & l^2n & l^2m & lu & lv \\
m^2n & lmn  & lm^2 & mu & mv \\
mn^2 & ln^2 & lmn  & nu & nv \\
\end{bmatrix},
\label{eqpdorbs}
\end{eqnarray}
where $\beta_{pd}(d)=\left\langle p_i^\pi\left|\nu_2\right|d_i^\pi\right\rangle=\beta_{pd}^{(0)}+\beta_{pd}^{(1)}\frac{d-d_0}{d_0}$ and
$\gamma_{pd}(d)=\sqrt{3}\left\langle p_i^\sigma\left|\nu_2\right|d_i^\sigma\right\rangle-2\left\langle p_i^\pi\left|\nu_2\right|d_i^\pi\right\rangle=\gamma_{pd}^{(0)}+\gamma_{pd}^{(1)}\frac{d-d_0}{d_0}$. This matrix is non zero in the unstrained diamond or zinc-blende crystal if $\gamma_{pd}^{(0)}\ne0$.

%\printtables
%\printfigures


\begin{thebibliography}{99}

\bibitem{ITRS} {\it International Technology Roadmap for Semiconductors}, available at http://www.itrs.net.
\bibitem{Thompson04} S.~E. Thompson, G. Sun, K. Wu, J. Lim, and T. Nishida, IEDM Tech. Digest 2004, 221 (2004).
\bibitem{Irie04} H. Irie, K. Kita, K. Kyuno, and A. Toriumi, IEDM Tech. Digest 2004, 225 (2004).
\bibitem{Payet06} F. Payet, F. Boeuf, C. Ortolland, and T. Skotnicki, SSDM Tech. Digest 2006, 176 (2006).
\bibitem{Schaffler97} F. Schaffler, Semicond. Sci. Technol. {\bf 12}, 1515 (1997).

\bibitem{Hohenberg64} P. Hohenberg and W. Kohn, Phys. Rev. {\bf 136}, B864 (1964).
\bibitem{Parr89} R.~G. Parr and W. Yang, {\it Density Functional Theory of Atoms and Molecules} (Oxford Universiy Press, New-York, 1989).
\bibitem{Walle86} C.~G. Van de Walle and R. M. Martin Phys. Rev. B {\bf 34}, 5621 (1986).
\bibitem{Cardona87} M. Cardona and N.~E. Christensen, Phys. Rev. B {\bf 35}, 6182 (1987).
\bibitem{Walle89} C.~G. Van de Walle, Phys. Rev. B {\bf 39}, 1871 (1989).
\bibitem{Resta90} R. Resta, L. Colombo, and S. Baroni, Phys. Rev. B {\bf 41}, 12358 (1990).
\bibitem{Wei99} S.-H Wei and A. Zunger, Phys. Rev. B {\bf 60}, 5404 (1999).
\bibitem{Li06} Y.-H. Li, X.~G. Gong, and S.-H. Wei, Phys. Rev. B {\bf 73}, 245206 (2006).

\bibitem{Bastard88} G. Bastard, {\it Wave Mechanics Applied to Semiconductor Heterostructures} (Les Editions de Physique, Les Ulis, 1988).
\bibitem{Rideau06} D. Rideau, M. Feraille, L. Ciampolini, M. Minondo, C. Tavernier, H. Jaouen, and A. Ghetti, Phys. Rev. B {\bf 74}, 195208 (2006).

\bibitem{Chelikowsky76} J.~R. Chelikowsky and M.~L. Cohen, Phys. Rev. B {\bf 14}, 556 (1976); Phys. Rev. B {\bf 30}, 4828(E) (1984).
\bibitem{Zunger95} L.-W. Wang and A. Zunger, Phys. Rev. B {\bf 51}, 17398 (1995).
\bibitem{Zunger99} L.-W. Wang and A. Zunger, Phys. Rev. B {\bf 59}, 15806 (1999).

\bibitem{Slater54} J.~C. Slater and G.~F. Koster, Phys. Rev. {\bf 94}, 1498 (1954).
\bibitem{Delerue05} C. Delerue and M. Lannoo, {\it Nanostructures: Theory and Modelling} (Springer, New York, 2004).
\bibitem{Niquet08} Y.~M. Niquet and D. Camacho Mojica, Phys. Rev. B {\bf 77}, 115316 (2008).
\bibitem{Klimeck07} G. Klimeck, S.~S. Ahmed, Hansang Bae, N. Kharche, S. Clark, B. Haley, Sunhee Lee, M. Naumov, Hoon Ryu, F. Saied, M. Prada, M. Korkusinski, T.~B. Boykin, and R. Rahman, IEEE Trans. on Electron Devices {\bf 54}, 2079 (2007); G. Klimeck, S.~S. Ahmed, N. Kharche, M. Korkusinski, M. Usman, M. Prada, and T.~B. Boykin, IEEE Trans. on Electron Devices {\bf 54}, 2090 (2007). 
\bibitem{Luisier06} M. Luisier, A. Schenk, W. Fichtner, and G. Klimeck, Phys. Rev. B {\bf 74}, 205323 (2006)
\bibitem{Luisier07} M. Luisier, A. Schenk, and W. Fichtner, Appl. Phys. Lett. {\bf 90}, 102103 (2007).
\bibitem{Svizhenko07} A. Svizhenko, P.~W.Leu, and K. Cho, Phys. Rev. B {\bf 75}, 125417 (2007).
\bibitem{Lherbier08} A. Lherbier, M.~P. Persson, Y.~M. Niquet, F. Triozon, and S. Roche, Phys. Rev. B {\bf 77}, 085301 (2008).
\bibitem{Martins05}  A.~S. Martins, T.~B. Boykin, G. Klimeck, and B. Koiller, Phys. Rev. B {\bf 72}, 193204 (2005).
\bibitem{Diarra08} M. Diarra, C. Delerue, Y.~M. Niquet, and G. Allan, J. Appl. Phys. {\bf 103}, 073703 (2008).
\bibitem{Delerue01} C. Delerue, G. Allan, and M. Lannoo, Phys. Rev. B {\bf 64}, 193402 (2001).
\bibitem{Jancu98} J.-M. Jancu, R. Scholz, F. Beltram, and F. Bassani, Phys. Rev. B {\bf 57}, 6493 (1998).
\bibitem{Harrison79} S. Froyen and W.~A. Harrison, Phys. Rev. B {\bf 20}, 2420 (1979).
\bibitem{Harrison80} W.~A. Harrison, {\it Electronic Structure and the Properties of Solids} (Freeman, San Francisco, 1980).
\bibitem{Priester88} C. Priester, G. Allan, and M. Lannoo, Phys. Rev. B {\bf 37}, 8519 (1988).
\bibitem{Brey82} L. Brey, C. Tejedor, and J.~A. Verg\'es, Phys. Rev. B {\bf 29}, 6840 (1984).
\bibitem{Tserbak93} C. Tserbak, H.~M. Polatoglou, and G. Theodorou, Phys. Rev. B {\bf 47}, 7104 (1993).
\bibitem{Boykin02} T.~B. Boykin, G. Klimeck, R.~C. Bowen, and F. Oyafuso, Phys. Rev. B {\bf 66}, 125207 (2002).
\bibitem{Boykin07} T.~B. Boykin, N. Kharche, and G. Klimeck, Phys. Rev. B {\bf 76}, 035310 (2007).
\bibitem{Jancu07} J.-M. Jancu and P. Voisin, Phys. Rev. B {\bf 76}, 115202 (2007).
\bibitem{noteavg} This implies the choice of a particular energy reference, see Refs. \onlinecite{Walle86,Cardona87,Walle89,Resta90,Wei99,Li06}.
%\bibitem{Pryor98} C. Pryor, J. Kim, L.~W. Wang, A.~J. Williamson, and A. Zunger, J. Appl. Phys. {\bf 83}, 2548 (1998).
\bibitem{Chadi89} D. Chadi, in {\it Atomistic Simulation of Materials Beyond Pair Potentials}, edited by V. Vitek and D. Srolovitz (Plenum Press, New-York, 1989).
\bibitem{Mercer94} J. L. Mercer Jr. and M. Y. Chou, Phys. Rev. B {\bf 49}, 8506 (1994).

\bibitem{notealpha} $\alpha_p$ should also include a contribution $\alpha_p^\prime$ from the on-site potential $V_1$ in Eq. (\ref{eqVp}), whose depth increases, in a first approximation, proportional to the tensile hydrostatic strain on the atom. The absolute valence band deformation potential $a_v$ can not, indeed, be reproduced without this central correction, as shown in semi-empirical pseudopotential theory [see, e.g., discussion in T. Mattila, L.-W. Wang and A. Zunger, Phys. Rev. B {\bf 59}, 15270 (1999)]. $\alpha_p^\prime$ is expected negative, at variance with the nearest neighbor contributions defined in paragraph \ref{subsectionp}.

\bibitem{notebeta} Since $\beta(d)$ and $\gamma(d)$ tend to zero when $d\to0$ or $d\to\infty$ (whatever the orbitals), they have at least one extremum in between. This has been confirmed using for example SIESTA orbitals and the screened pseudopotential for Si from L.-W. Wang and A. Zunger, J. Chem. Phys. {\bf 100}, 2394 (1994). The sign of the first-order $\beta^{(1)}$ and $\gamma^{(1)}$ is therefore expected to be dependent on the parametrization, especially in an orthogonal approximation where the orbitals have long-range oscillations.

\bibitem{SIESTA} J.-M. Soler, E. Artacho, J. D. Gale, A. Garc\'ia, J. Junquera, P. Ordej\'on, and D. S\'anchez-Portal, J. Phys.: Condens. Matter {\bf 14}, 2745 (2002).

\bibitem{Jancu02} J.-M. Jancu, F. Bassani, F. Della Sala, and R. Scholz, Appl. Phys. Lett. {\bf 81}, 4838 (2002).

\bibitem{Keating66} P. N. Keating, Phys. Rev. {\bf 145}, 637 (1966).
\bibitem{Stillinger85} F.~H. Stillinger and T.~A. Weber, Phys. Rev. B {\bf 31}, 5262 (1985).
\bibitem{Kleinman62} L. Kleinman, Phys. Rev. {\bf 128}, 2614 (1962).

\bibitem{abinit} The ABINIT code is a common project of the Universit\'e Catholique de Louvain, Corning Incorporated, the Ecole Polytechnique of Palaiseau and other contributors (see http://www.abinit.org).
\bibitem{abinit02} X. Gonze, J.-M. Beuken, R. Caracas, F. Detraux, M. Fuchs, G.-M. Rignanese, L. Sindic, M. Verstraete, G. Zerah, F. Jollet, M. Torrent, A. Roy, M. Mikami, Ph. Ghosez, J.-Y. Raty, and D.~C. Allan, Computational Materials Science {\bf 25}, 478 (2002). 
\bibitem{abinit05} X. Gonze, G.-M. Rignanese, M. Verstraete, J.-M. Beuken, Y. Pouillon, R. Caracas, F. Jollet, M. Torrent, G. Zerah, M. Mikami, Ph. Ghosez, M. Veithen, J.-Y. Raty, V. Olevano, F. Bruneval, L. Reining, R. Godby, G. Onida, D.~R. Hamann, and D.~C. Allan, Zeit. Kristallogr. {\bf 220}, 558 (2005).
\bibitem{Hartwigsen98} C. Hartwigsen, S. Goedecker, and J. Hutter, Phys. Rev. B {\bf 58}, 3641 (1998).
\bibitem{Hedin65} L. Hedin, Phys Rev. {\bf 139}, A796 (1965).
\bibitem{Aulbur00} W. G. Aulbur, L. Jonsson, and J. W. Wilkins, Solid State Phys. {\bf 54}, 1 (2000).
\bibitem{detailonGW} A detailed description of the present GW results can be found in Ref. \onlinecite{Rideau06}. In particular, as mentioned in Ref. \onlinecite{Rideau06}, LDA-G$_0$W$_0$ results do not match {\it perfectly} the experimental data, and a supplementary rigid "scissor" shift of 0.09 eV for Si and 0.104 eV for Ge as been added to the GW results to obtain the final reference set of bands.
\bibitem{detailonGWstr} The same GW correction as in the bulk, unstrained materials is used in strained Si and Ge. This choice is motivated by X. Zhu, S. Fahy, and S. G. Louie, Phys. Rev. B {\bf 39}, 7840 (1989), who reported that there is no quantitative differences between the LDA and GW band gap pressure dependencies in Si.
\bibitem{Jones93} D.~R. Jones, C.~D. Perttunen, and B.~E. Stuckman, J. Optim. Theory App. {\bf 79}, 157 (1993).
\bibitem{Byrd93} R.~H. Byrd, P. Lu, and J. Nocedal, SIAM Journal on Scientific and Statistical Computing {\bf 16}, 1190 (1995).
\bibitem{notefithydro} We actually fit the tight-binding parameters on the differences between the strained and unstrained band structures, in order to achieve a good description of the deformation potentials, free from the errors on the unstrained band structure.
\bibitem{notegamma0} We have, in particular, attempted to optimize simultaneously the on-site energies, nearest neighbor interactions, $\beta^{(0)}$'s and $\gamma^{(0)}$'s on the unstrained band structure, and on one $[001]$ and one $[111]$ biaxial strain that do not change the nearest neighbor bond lengths (to be independent on the Harrison parameters, $\alpha$'s, $\beta^{(1)}$'s and $\gamma^{(1)}$'s). We did not evidence any real improvement in the description nor any change in the signs of the $\beta^{(0)}$'s.

\bibitem{Landolt} {\it Physics of Group IV elements and III-V Compounds}, edited by O. Madelung, M. Schulz, and H. Weiss, Landolt-B\"ornstein, New series, Group III, Vol. 17, Pt. A (Springer-Verlag, New-York, 1982).
\bibitem{Humphreys81} R.~G. Humphreys, J. Phys. C: Solid State Phys. {\bf 14}, 2935 (1981).
\bibitem{Lawaetz71} P. Lawaetz, Phys. Rev. B {\bf 4}, 3460 (1971).
\bibitem{Levinger61} B.~W. Levinger and D.~R. Frankl, J. Phys. Chem. Solids {\bf 18}, 139 (1961).
\bibitem{Halpern65} J. Halpern and B. Lax, J. Phys. Chem. Solids {\bf 26}, 911 (1965).
\bibitem{Dresselhaus55} G. Dresselhaus, A.~F. Kip, and C. Kittel, Phys. Rev. {\bf 98}, 368 (1955).
\bibitem{Makarov83} O.~A. Makarov, N.~N. Ovsyuk, and M.~P. Sinyukov, Sov. Phys. JETP {\bf 57}, 1318 (1983).
\bibitem{Boykin04} T.~B. Boykin, G. Klimeck, and F. Oyafuso, Phys. Rev. B {\bf 69}, 115201 (2004).

\bibitem{Colombo91} L. Colombo, R. Resta, and S. Baroni, Phys. Rev. B {\bf 44}, 5572 (1991).
\bibitem{Schwartz89} G.~P. Schwartz, M.~S. Hybertsen, J. Bevk, R.~G. Nuzzo, J.~P. Mannaerts, and G.~J. Gualtieri, Phys. Rev. B {\bf 39}, 1235 (1989).

\bibitem{noteuniax} To plot Figs. \ref{figUniaxSi} and \ref{figUniaxGe}, we used a constant $\varepsilon_\perp/\varepsilon_\parallel$ ratio and the {\it ab initio} internal strain parameter $\zeta$ fitted to a second-order polynomial: $\varepsilon_\perp/\varepsilon_\parallel=-0.7708$ for Si $[001]$, $\varepsilon_\perp/\varepsilon_\parallel=-0.4392$ for Si $[111]$ ($\zeta=0.5213-1.1156\varepsilon_\parallel+3.1868\varepsilon_\parallel^2$), $\varepsilon_\perp/\varepsilon_\parallel=-0.51$ for Si $[110]$ ($\zeta=0.5254-2.9765\varepsilon_\parallel+3.4661\varepsilon_\parallel^2$); $\varepsilon_\perp/\varepsilon_\parallel=-0.7518$ for Ge $[001]$, $\varepsilon_\perp/\varepsilon_\parallel=-0.3717$ for Ge $[111]$ ($\zeta=0.4813-0.9972\varepsilon_\parallel+0.6567\varepsilon_\parallel^2$), and $\varepsilon_\perp/\varepsilon_\parallel=-0.4504$ for Ge $[110]$ ($\zeta=0.4783-2.8113\varepsilon_\parallel+4.1082\varepsilon_\parallel^2$).
\bibitem{noteavlda} Since {\it ab initio} calculations performed at different strains do not give the band structure on an {\it absolute}, common energy scale, we choose to align the {\it ab initio} and TB highest valence bands on Figs. \ref{figUniaxSi} and \ref{figUniaxGe}. Therefore, only the other bands are indicative of the quality of the TB model.
\bibitem{Ungersboeck07} E. Ungersboeck, S. Dhar, G. Karlowatz, V. Sverdlov, H. Kosina, and S. Selberherr, IEEE Trans. on Electron Devices {\bf 54}, 2183 (2007).
\bibitem{Yamauchi08} J. Yamauchi, IEEE Electron Device Lett. {\bf 29}, 186 (2008).

\bibitem{Weber89} J. Weber and M.~I. Alonso, Phys. Rev. B {\bf 40}, 5683 (1989).
\bibitem{Dismukes64} J.~P. Dismukes, L. Ekstrom, and R.~J. Paff, J. Phys. Chem. {\bf 68}, 3021 (1964).
%\bibitem{Dutartre91} D. Dutartre, G. Br\'emond, A. Souifi and T. Benyattou, Phys. Rev. B {\bf 44}, 11525 (1991). % 001.
\bibitem{Robbins92} D.~J. Robbins, L.~T. Canham, S.~J. Barnett, A.~D. Pitt, and P. Calcott, J. Appl. Phys. {\bf 71}, 1407 (1992).
\bibitem{Lang85} D.~V. Lang, R. People, J.~C. Bean, and A.~M. Sergent, Appl. Phys. Lett. {\bf 47}, 1333 (1985).
\bibitem{Spitzer92} J. Spitzer, K. Thonke, R. Sauer, H. Kibbel, H.~J. Herzog, and E. Kasper, Appl. Phys. Lett. {\bf 60}, 1729 (1992).
\bibitem{Liu94} C.~W. Liu, J.~C. Sturm, Y.~R.~J. Lacroix, M.~L.~W. Thewalt, and D.~D. Perovic, Appl. Phys. Lett. {\bf 65}, 76 (1994). % 110.
\bibitem{Takagi98} S. Takagi, J.~L. Hoyt, K. Rim, J.~J. Wesler, and J.~F. Gibbons, IEEE Trans. on Electron Devices {\bf 45}, 494 (1998).
\bibitem{Fischetti88} M.~V. Fischetti and S.~E. Laux, Phys. Rev. B {\bf 38}, 9721 (1988).
\bibitem{Pham07} Anh-Tuan Pham, C. Jungemann, and B. Meinerzhagen, IEEE Trans. on Electron Devices {\bf 54}, 2174 (2007).

% Appendix.

\bibitem{notecfp} These twofold and threefold degenerate basis functions will be appropriate linear combinations of the $d$ orbitals for other orientations of the crystal with respect to the $x$, $y$ and $z$ axes.

\end{thebibliography}
\end{document}